%% file: main.tex
\documentclass[preprint]{aastex61}

\usepackage[]{natbib}
\usepackage{booktabs, graphicx, multirow, array, amsmath}

\def\arcsec{$^{\prime\prime}$}

\def\nnh{N${_2}$H$^{+}$}
\def\hco{HCO$^{+}$}
\def\htco{H$^{13}$CO$^{+}$}
\def\htcn{H$^{13}$CN}
\def\deg{$^{\circ}$}
\newcommand{\Msun}{\mbox{$M_{\sun}$}}

\begin{document}

\title{Morphology and Kinematics of Filaments in the Serpens and Perseus Molecular Clouds}

\author{Arnab Dhabal}
\affiliation{Department of Astronomy, University of Maryland, College Park, MD 20742, USA}
\affiliation{NASA Goddard Space Flight Center, Greenbelt, MD 20771, USA}
\email{adhabal@astro.umd.edu}

\author{Lee G. Mundy}
\affiliation{Department of Astronomy, University of Maryland, College Park, MD 20742, USA}
\email{lgm@astro.umd.edu}

\author{Maxime J. Rizzo}
\affiliation{Department of Astronomy, University of Maryland, College Park, MD 20742, USA}
\affiliation{NASA Goddard Space Flight Center, Greenbelt, MD 20771, USA}

\author{Shaye Storm}
\affiliation{Department of Astronomy, University of Maryland, College Park, MD 20742, USA}
\affiliation{Harvard-Smithsonian Center for Astrophysics, 60 Garden Street, Cambridge, MA 02138, USA}

\author{Peter Teuben}
\affil{Department of Astronomy, University of Maryland, College Park, MD 20742, USA}

\begin{abstract}
We present \htco\ (J=1-0) and HNC (J=1-0) maps of regions in Serpens South, Serpens Main and NGC 1333 containing filaments. We also observe the Serpens regions using \htcn\ (J=1-0). These dense gas tracer molecular line observations carried out with CARMA have an angular resolution of $\sim$ 7\arcsec, a spectral resolution of $\sim$ 0.16 $km/s$ and a sensitivity of 50-100 $mJy/beam$. Although the large scale structure compares well with the \textit{Herschel} dust continuum maps, we resolve finer structure within the filaments identified by \textit{Herschel}. The \htco\ emission distribution agrees with the existing CARMA \nnh\ (J=1-0) maps; so they trace the same morphology and kinematics of the filaments. The \htco\ maps additionally reveal that many regions have multiple structures partially overlapping in the line-of-sight. In two regions, the velocity differences are as high as 1.4 $km/s$. We identify 8 filamentary structures having typical widths of $0.03-0.08$ $pc$ in these tracers. At least 50\% of the filamentary structures have distinct velocity gradients perpendicular to their major axis with average values in the range $4-10$ $km\ s^{-1} pc^{-1}$. These findings are in support of the theoretical models of filament formation by 2-D inflow in the shock layer created by colliding turbulent cells. We also find evidence of velocity gradients along the length of two filamentary structures; the gradients suggest that these filaments are inflowing towards the cloud core.
\end{abstract}


\input{intro.tex}
\input{observations.tex}

\input{results.tex}
\input{analysis.tex}
\input{discussion.tex}
\input{summary.tex}

\begin{acknowledgments}
The authors would like to thank all members of the CARMA staff who made these observations possible. The CARMA operations were funded by the National Science Foundation and the consortium universities. The CLASSy project was supported by AST-1139950 (University of Illinois) and AST-1139998 (University of Maryland). This research was also supported through the NASA Grant NNX11AG92A.
\end{acknowledgments}

\software{MIRIAD \citep{Sault1995}, Astropy \citep{Astropy2013},
          Matplotlib \citep{Matplotlib2007}}

\bibliographystyle{aasjournal}
\bibliography{refCARMAfil}

\end{document}

%% file: intro.tex
\section{Introduction}
\label{sec:Intro}

The presence of filaments in molecular clouds has been well known for many decades \citep{SchE1979, Bally1987, Chini1997, Goldsmith2008}. Far infrared continuum images taken by \textit{Herschel} demonstrated the prevalence of filamentary structure over a wide range of scales, and trace filaments into areas of active star formation \citep{Andre2010, Molinari2010}. There is growing observational evidence that filaments play a fundamental role in the star formation process by setting the initial conditions for core formation and fragmentation, and defining the morphology of the material available for accretion \citep{Hacar2011, Furuya2014, Beu2015, Friesen2016, Henshaw2016, Teix2016}. Some cores within filaments evolve to harbor YSOs, the properties of which depend on the filament physical characteristics \citep{Myers2009, Kritsuk2013}. 

Filaments are also commonly present in numerical simulations of turbulent molecular gas \citep{Klessen2000, Boldyrev2002, Banerjee2006, Hei2009, GongOst2011, Peters2012, GVS2014}. These filaments form at the interface of converging flows in regions of supersonic turbulence. However the impacts of self-gravity \citep{Moeckel2015}, turbulence \citep{Smith2014} and magnetic fields \citep{Kirk2015} on filament formation and structure are still under study. Filament morphologies, their widths, densities, orientations with each other and with respect to local magnetic fields, can provide important insights into the filament formation process. For example, the systematics of filament width could be related to the sonic scale of turbulence \citep{Fed2016}, ion-neutral friction \citep{Hennb2013}, or the presence of magnetic fields \citep{SW2015}. Maps using molecular tracers with high angular resolution and/or high spectral resolution resolved some filamentary structures further to multiple filaments \citep{Lee2014, Hacar2013}, which are also found in simulations \citep{Moeckel2015, Kirk2015, Smith2016, Clarke2017}.

The kinematics of filaments \citep{Sch2010, JS2014} can be useful indicators of the evolutionary processes involved in star formation. For example, velocity gradients along filaments have been postulated either as mass infalls towards the clumps \citep{Kirk2013}, as projections of large-scale turbulence \citep{FL2014}, or as convergence of multiple flows \citep{Henshaw2013, Csengeri2011}. On the other hand, a velocity gradient perpendicular to the major axis has been postulated as effects of filament rotation \citep{OT2002}, or infall due to self-gravity of gas compressed into a planar structures by supersonic turbulence \citep{Smith2016, GongOst2011}. Supersonic velocity dispersions are expected in the case of gravitationally accelerated infalling material which enter the filaments through shocks at the boundaries \citep{Hei2009}. 

\begin{figure*}
\includegraphics[width=\textwidth]{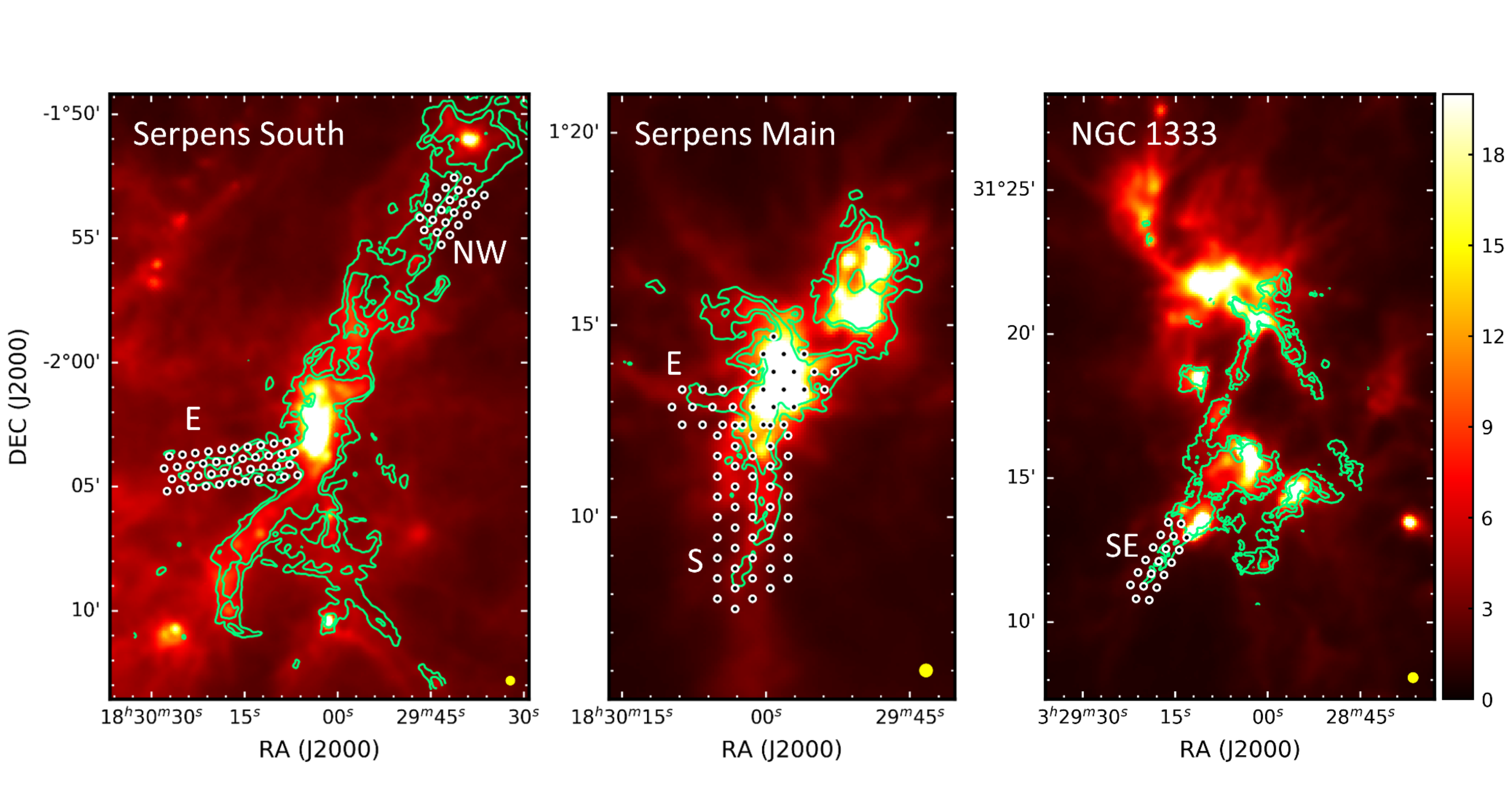}
\caption{\textit{Herschel} 250 $\mu m$ maps (Units: $Jy\ beam ^{-1}\ km\ s^{-1}$) overlaid with \nnh\ integrated intensity contours (green) from CLASSy observations of the Serpens South, Serpens Main and NGC 1333 molecular clouds. The contours are at (3,6) $\times$ $\sigma$, where $\sigma$ = 0.25, 0.7 and 0.4 $Jy\ beam ^{-1}\ km\ s^{-1}$ respectively for the three regions. The circles mark the CLASSy-II mosaic pointing centers and indicate the regions observed. The regions correspond to filaments that we name based on their locations with respect to the parent cloud cores. The \textit{Herschel} beam is 18\arcsec\ in diameter (shown in the bottom-right corner of each image).}
\label{fig:Herschel}
\end{figure*}

This paper presents high angular resolution maps of dense gas filaments in three active star formation regions --Serpens Main, Serpens South, and NGC 1333 to study the detailed morphology and kinematics of five filaments. This study derives from the CARMA Large Area Star Formation Survey (CLASSy), a CARMA key project which imaged 800 square arcminutes of Perseus and Serpens Molecular Clouds using the \nnh, HCN and \hco\ J=1-0 lines \citep{Storm2014,Lee2014, FL2014}. These clouds host a large number of young stars and protostars associated with hub-filament type gas structures \citep{Guter2008I, Guter2008II}, and are hence well-suited for studying the connection between filaments and star formation. \nnh\ emission, which is a cold, dense gas tracer \citep{Taf2002}, was found to closely follow the far-infrared filamentary structure in high column density regions \citep{Storm2014, Lee2014, FL2014}. Figure \ref{fig:Herschel} shows the \nnh\ emission overlaid on the \textit{Herschel} dust continuum emission at 250 $\mu$m. The \nnh\ emission also revealed distinctive velocity gradients in a number of filaments. The HCN and \hco\  emission, on the other hand, were found to be poor tracers of filaments. \cite{Storm2014} proposed that these lines were optically thick, and in the case of \hco\ affected by higher fractional abundances in lower density gas which create an overlying absorption region.

To investigate whether the kinematics of filamentary structures are accurately traced by the CLASSy \nnh\ observations, we selected 5 filaments from the CLASSy regions to be observed using optically thin dense gas tracers \htco\ and \htcn. The mapped regions are marked in Figure \ref{fig:Herschel} by the white circles. These molecules do not have the same chemistry as \nnh; hence they serve as a test of whether the observed kinematics is a bulk property of the material or if they arise from chemical or excitation effects. We refer to the observations presented in this paper as CLASSy-II observations.

The layout of the paper is as follows. In Section \ref{sec:Obs}, we discuss how the observations were carried out followed by the data calibration and reduction procedure. We present the maps for the various tracers for all the regions in Section \ref{sec:Res}. Section \ref{sec:Analysis} deals with analyzing the individual regions and characterizing them. In Section \ref{sec:Disc}, we report the trends in the morphology and kinematics of these observed regions. We further discuss some of these results in context of the major questions in filament formation and their relation to star formation. We summarize our main results in Section \ref{sec:Sum}.

%% file: observations.tex
\section{Observations}
\label{sec:Obs}

The observations were made using the full CARMA array of 23 antennas in D and E configurations, providing an angular resolution of 7\arcsec. In CARMA 23-element mode, the correlator has four bands. In addition to the $^{13}C$ isotopologues of \hco\ and HCN, HNC was also available in the correlator setting as a third tracer. In case of NGC 1333, \nnh\ was observed instead of \htcn. The fourth band was used to observe the continuum at 90 $GHz$ over a bandwidth of 500 $MHz$. The spectral line bands were 7.7 $MHz$ wide with 159 channels, providing a resolution of $\sim$ 0.16 $km/s$. The summary of the observations are given in the Table \ref{tbl:Obs}, while the properties of the molecular tracer transitions are given in Table \ref{tbl:Lines}. The observations were carried out for a total of 308 hours over 61 tracks between August 2013 and February 2015.

\begin{table*}[h!]
\centering
\input{Tbl_Regions.tex}
\caption{Summary of Observations}
\label{tbl:Obs}
\end{table*}

\begin{table*}
\centering
\input{Tbl_Lines.tex}
\caption{Observed Molecular Lines}
\label{tbl:Lines}
\end{table*}

Of the three molecular tracers, \htco\ J=1-0 has the simplest spectrum: a single line with no hyperfine components. \htcn\ J=1-0 has three isolated hyperfine components. HNC J=1-0 \citep{Bechtel2006} has hyperfine components within a range of 200 $kHz$ (0.66 km sec$^{-1}$), which are barely resolved by our correlator channel width of 49 $kHz$; the primary effect is to modestly increase the observed linewidth. For comparison, \nnh\ J=1-0 has seven hyperfine components, of which three $F_1$=1-1 lines are within a range of 600 $kHz$, three $F_1$=2-1 lines within 500 $kHz$, and one isolated $F_1$=0-1 line. Although all four molecules are dense gas tracers, \htcn\ has the highest critical density $\sim 10^6$ $cm^{-3}$, while the other transitions have critical densities in the range $1.4-2.8$ $\times 10^5$ $cm^{-3}$ (see Table \ref{tbl:Lines}).

The flux calibrators for individual tracks were MWC349, Mars, Mercury and Uranus, as available during observations. The phase calibrators were 3C84 and 3C111 for NGC 1333, and 1743-048 for all Serpens regions. Although a bandpass calibrator was observed in all observation sessions, in many cases the phase calibrators were also used for bandpass calibration. 

The \texttt{MIRIAD} package (Multichannel Image Reconstruction, Image Analysis and Display; \citep{Sault1995}) was used to process the visibility data. The autocorrelation data from the 10 m antennas were used to make single-dish images. After iterative flagging and calibration of the visibilities, the single dish maps were used with the interferometric visibility data to generate the data-cubes using the \texttt{MIRIAD} Maximum Entropy Method program \texttt{mosmem}. The combined deconvolution using the single dish images helps in recovering the large-scale structure filtered out by the interferometer. The spectral channel RMS noise is in the range of $50-125$ $mJy/beam$, while the continuum RMS noise is $0.4-1.0$ $mJy/beam$ for the various fields.

%% file: Tbl_Regions.tex
\newcolumntype{P}[1]{>{\centering\arraybackslash}p{#1}}

\begin{tabular}{p{12em} c P{3em} P{3em} P{5em} P{5.5em}} 
\toprule
Region Name & Lines observed & \multicolumn{2}{c}{Hours observed} & \multirow{2}{5em}{\centering{No. of pointings}} & \multirow{2}{6em}{\centering{Synthesized Beam Size*}}\\ \cline{3-4}
& & D & E & &\\ 
\midrule 
Serpens South - NW region & H$^{13}$CO$^{+}$, HNC, H$^{13}$CN & 51.9 & 41 & 22 & 9.1" $\times$ 6.8"\\
Serpens South - E region & H$^{13}$CO$^{+}$, HNC, H$^{13}$CN & 31.3 & 20.6 & 42 & 8.9" $\times$ 6.0"\\
Serpens Main - cloud center and E filament & H$^{13}$CO$^{+}$, HNC, H$^{13}$CN & 18.6 & 43.6 & 30 & 8.4" $\times$ 7.1"\\
Serpens Main - S region & H$^{13}$CO$^{+}$, HNC, H$^{13}$CN & 24.1 & 19.4 & 45 & 9.5" $\times$ 6.4"\\
NGC 1333 - SE region & H$^{13}$CO$^{+}$, HNC, N$_2$H$^{+}$ & 28.7	& 29 & 19 & 7.2" $\times$ 5.4"\\
\bottomrule
\end{tabular}

\raggedright{\footnotesize{*Final synthesized beam sizes obtained after Maximum Entropy deconvolution}}

%% file: Tbl_Lines.tex
\begin{tabular}{ c p{5em} p{5em} p{5em}} 
\toprule
Lines & Frequency (GHz) & $n_{cr}$@10K ($cm^{-3}$)$^a$ & Velocity Resolution$^b$ (km/s)\\ 
\midrule 

H$^{13}$CO$^{+}$ J=1-0 & 86.75429 & 1.5 $\times$ 10$^{5}$ & 0.169\\
H$^{13}$CN J=1-0 & 86.33986 & 2.0 $\times$ 10$^{6}$ & 0.170\\
HNC J=1-0 & 90.66357 & 2.8 $\times$ 10$^{5}$ & 0.162\\
N$_2$H$^{+}$ J=1-0 & 93.17370 &  1.4 $\times$ 10$^{5}$ & 0.158\\
\bottomrule
\end{tabular}

\raggedright{\footnotesize{$^a$ Critical densities ($n_{cr}$) are calculated from the ratio of Einstein A coefficient values $A_{ul}$ and collision rate coefficients $\gamma_{ul}$. These values are obtained from the Leiden molecular database \citep{Leiden2005} and are based on $\gamma_{ul}$ values from \cite{Flower1999}, \cite{Green1974} and \cite{Dumouchel2010}. \newline$^b$ Calculated based on channel widths of 49 $kHz$.}}

%% file: results.tex
\section{Results}
\label{sec:Res}

We generated integrated intensity (0th Moment) and velocity (1st Moment) maps from the position-velocity data cubes using \texttt{MIRIAD}. We used only the channels near the line centers containing the signal, and clipped the data at 2$\sigma$, based on the RMS noise of each line map. For the first moment maps of \nnh\ and \htcn, line fitting including the contribution of all the hyperfine components was carried out to determine the velocity centroid.

\begin{figure*}
\includegraphics[trim={1.5cm 1cm 2.2cm 1.75cm},clip,width=\textwidth]{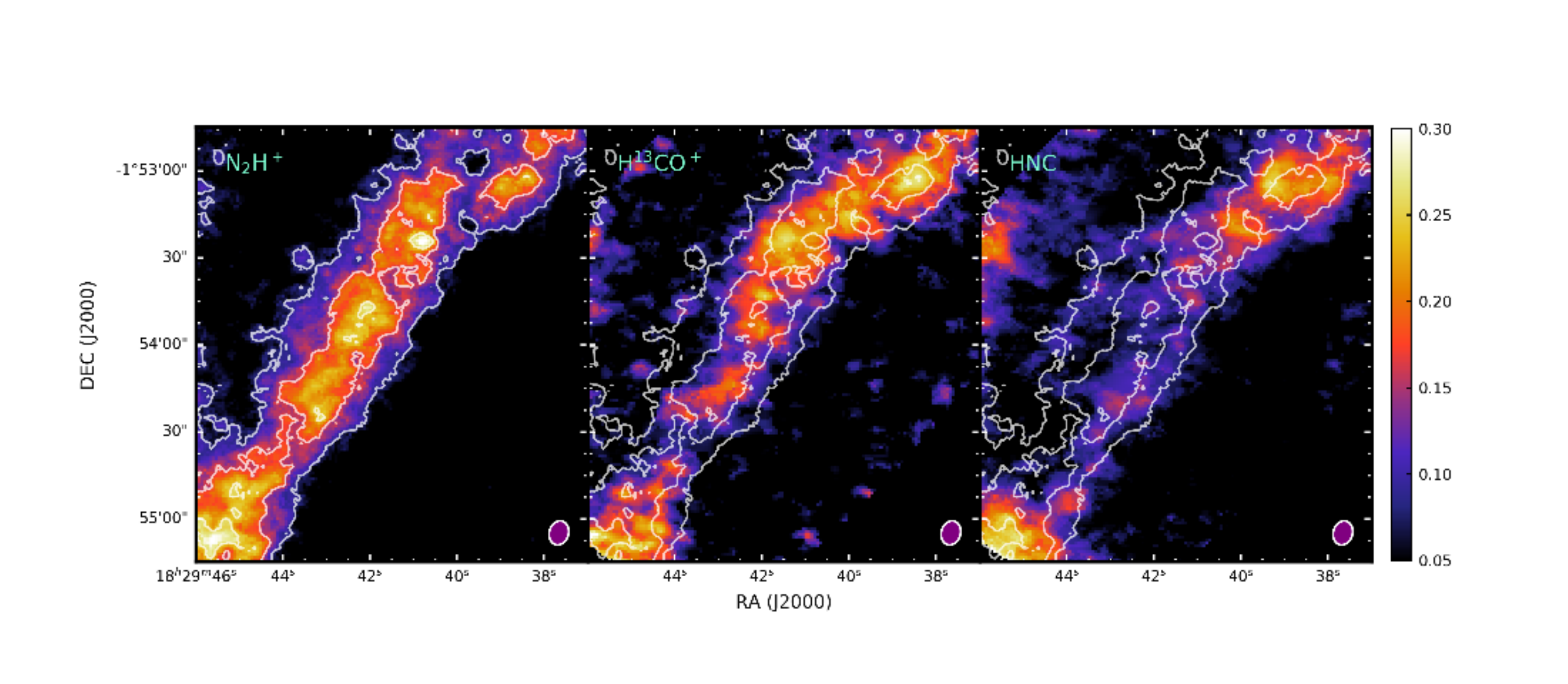}
\includegraphics[trim={1.5cm 1cm 2.2cm 1.75cm}, clip, width=\textwidth]{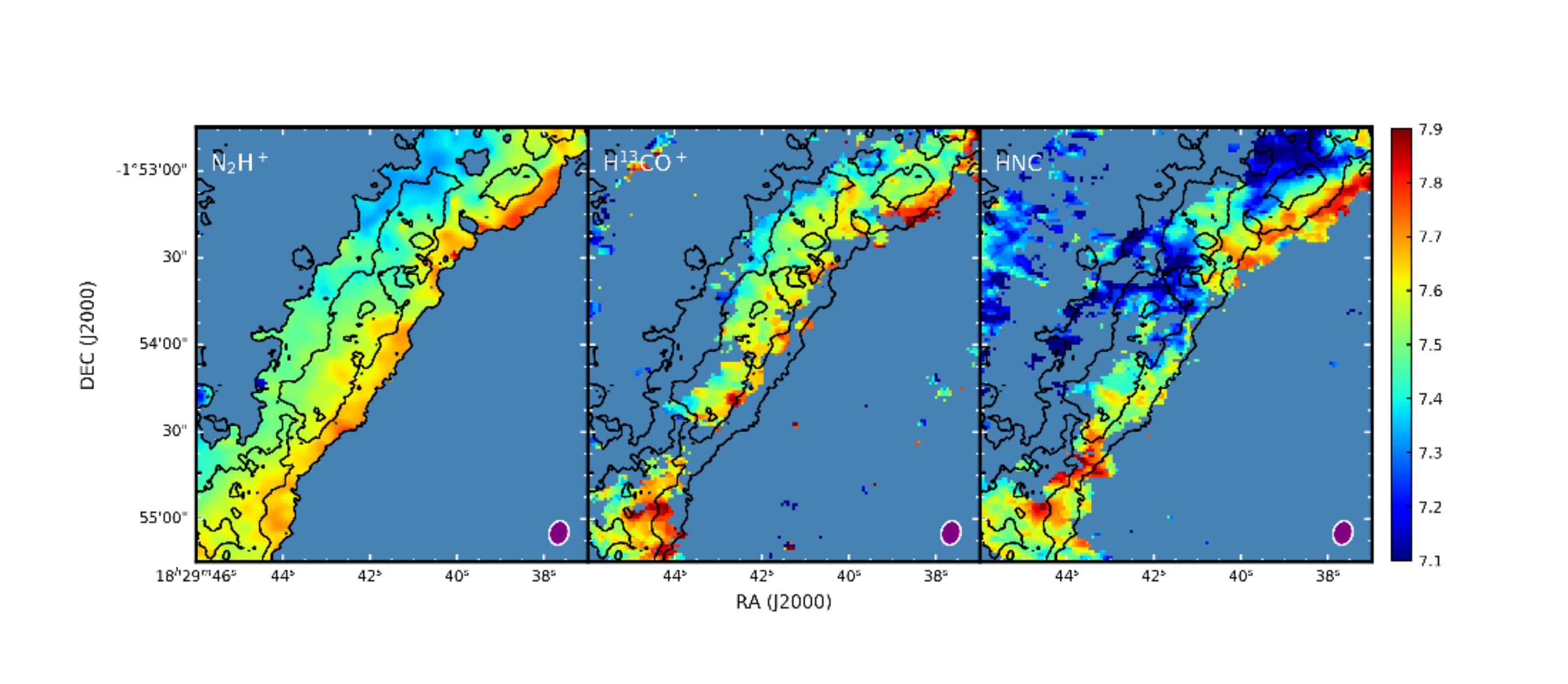}
\caption{Integrated intensity maps (top row) and velocity centroid maps (bottom row) for Serpens South - NW filament in the \nnh, \htco\ and HNC J=1-0 lines. The integrated intensity map color bar values are given for \htco\ in $Jy\ beam ^{-1}\ km\ s^{-1}$. The corresponding values for \nnh\ and HNC maps are respectively 6.0 and 3.0 times the value for the \htco\ maps. The velocity maps are all in units of $km\ s^{-1}$. The \nnh\ integrated intensity contours at (2,4,6) $\times$ $\sigma$ ($\sigma$ = 0.25 $Jy\ beam ^{-1}\ km\ s^{-1}$) are overlaid on all the images. The synthesized beam is shown at the bottom-right of each image.}
\label{fig:SERPSNW}
\end{figure*}

\begin{figure*}
\includegraphics[trim={2cm 0cm 2.2cm 0.5cm},clip,width=\textwidth]{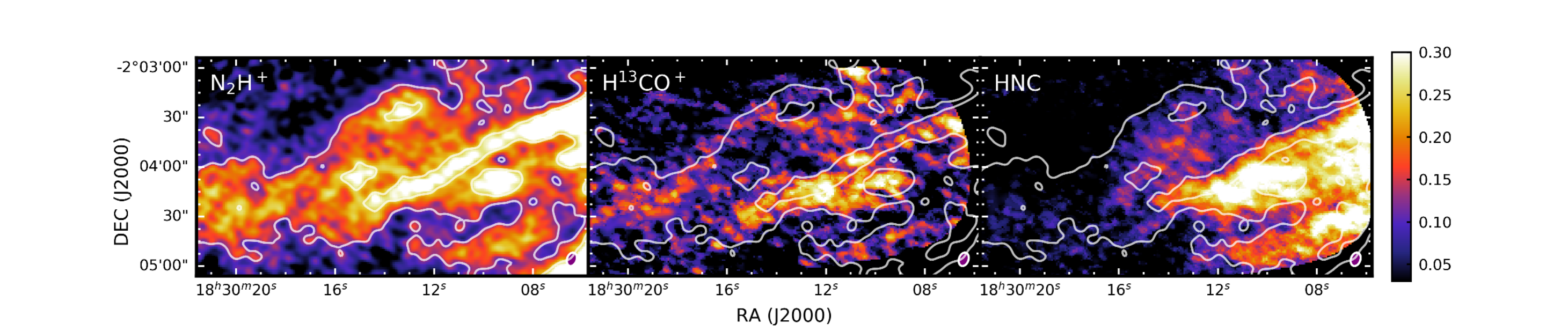}
\includegraphics[trim={2cm 0cm 2.2cm 0.5cm},clip,width=\textwidth]{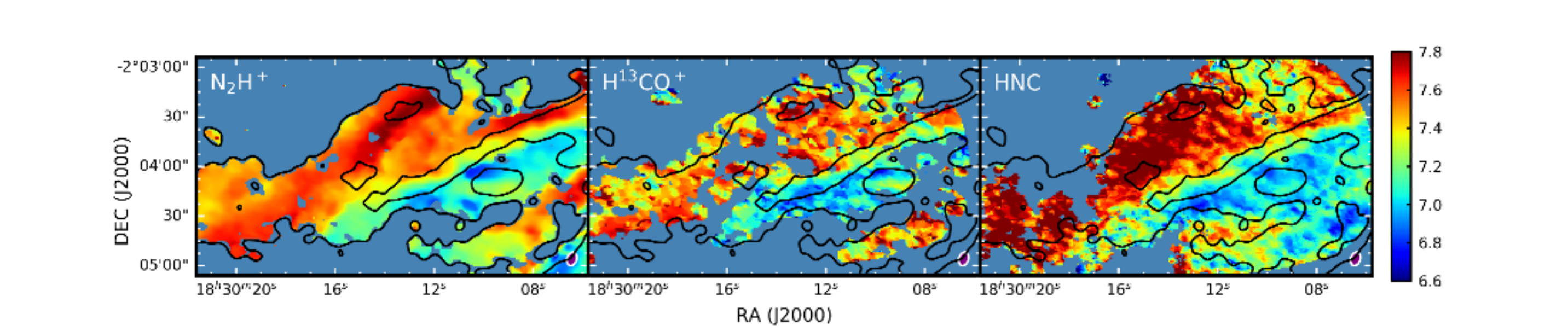}
\caption{Same as Fig. \ref{fig:SERPSNW} for Serpens South - E region. The integrated intensity maps are in units of 10.0, 1.0 and 4.0  $Jy\ beam ^{-1}\ km\ s^{-1}$ for \nnh, \htco, and HNC respectively. The \nnh\ contours are at (5,10) $\times$ $\sigma$ ($\sigma$ = 0.25 $Jy\ beam ^{-1}\ km\ s^{-1}$).}
\label{fig:SERPSE}
\end{figure*}

\begin{figure*}
\includegraphics[trim={1.5cm 0cm 2.2cm 0.5cm},clip,width=\textwidth]{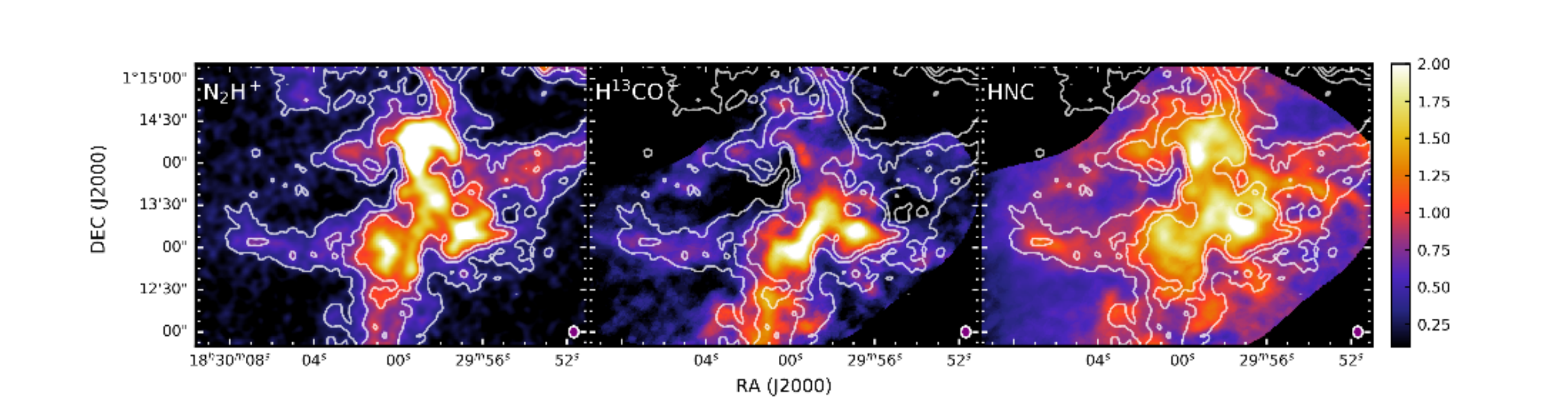}
\includegraphics[trim={1.5cm 0cm 2.2cm 0.5cm},clip,width=\textwidth]{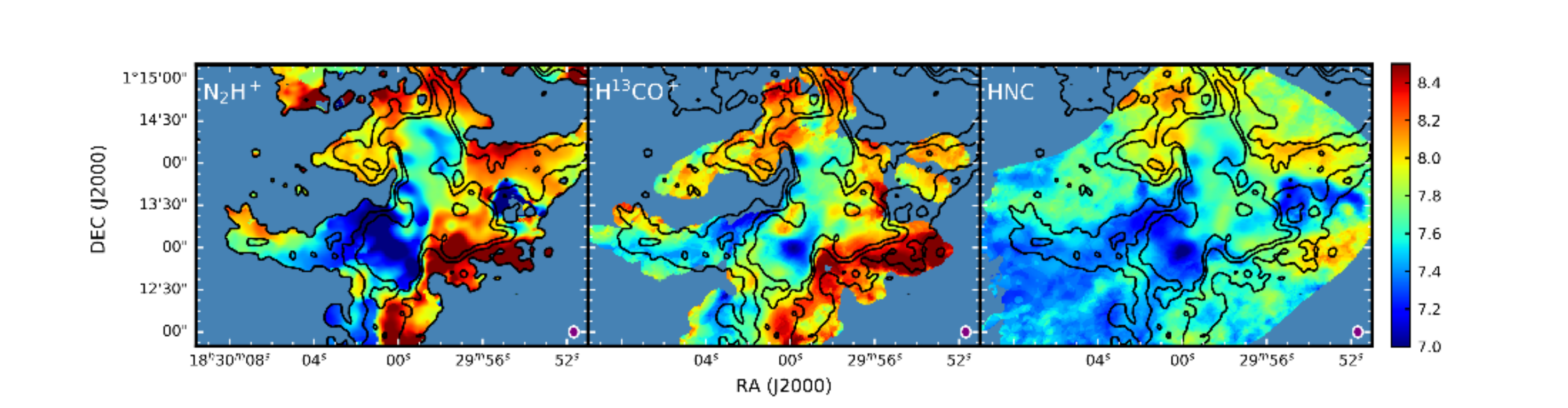}
\caption{Same as Fig. \ref{fig:SERPSNW} for Serpens Main - cloud center and E filament. The integrated intensity maps are in units of 6.0, 1.0 and 3.0  $Jy\ beam ^{-1}\ km\ s^{-1}$ for \nnh, \htco, and HNC respectively. The \nnh\ contours are at (3,6,9) $\times$ $\sigma$ ($\sigma$ = 0.7 $Jy\ beam ^{-1}\ km\ s^{-1}$).}
\label{fig:SERPME}
\end{figure*}

\begin{figure*}
\centering
\includegraphics[trim={0.5cm 1cm 1cm 2cm},clip,width=0.6\textwidth]{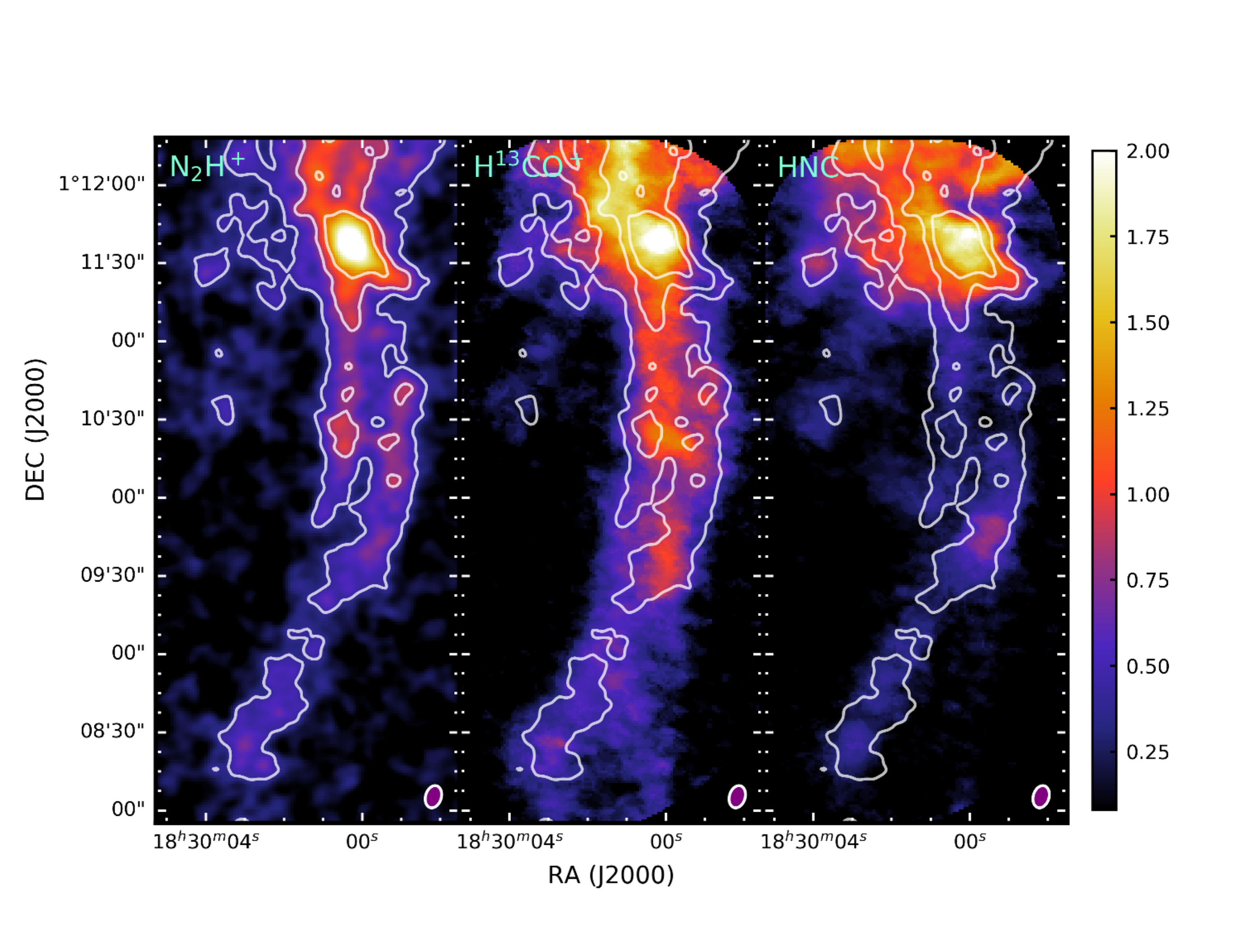} \break
\includegraphics[trim={0.5cm 1cm 1cm 2cm},clip,width=0.6\textwidth]{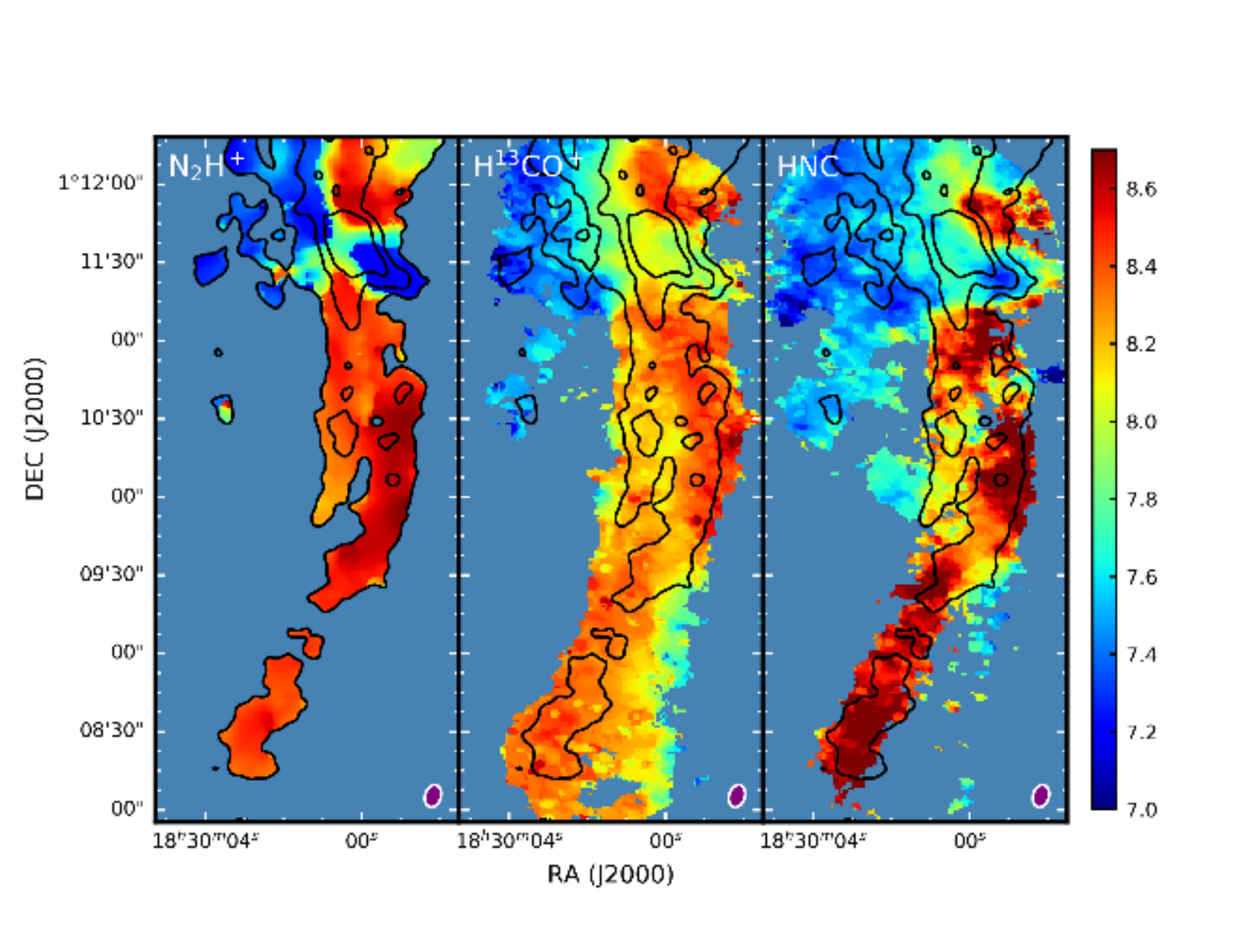}
\caption{Same as Fig. \ref{fig:SERPSNW} for Serpens Main - S region. The integrated intensity maps are in units of 5.0, 1.0 and 2.0  $Jy\ beam ^{-1}\ km\ s^{-1}$ for \nnh, \htco, and HNC respectively. The \nnh\ contours are at (3,6,9) $\times$ $\sigma$ ($\sigma$ = 0.7 $Jy\ beam ^{-1}\ km\ s^{-1}$).}
\label{fig:SERPMS}
\end{figure*}

\begin{figure*}
\includegraphics[trim={1.5cm 1cm 2.2cm 1.75cm},clip,width=\textwidth]{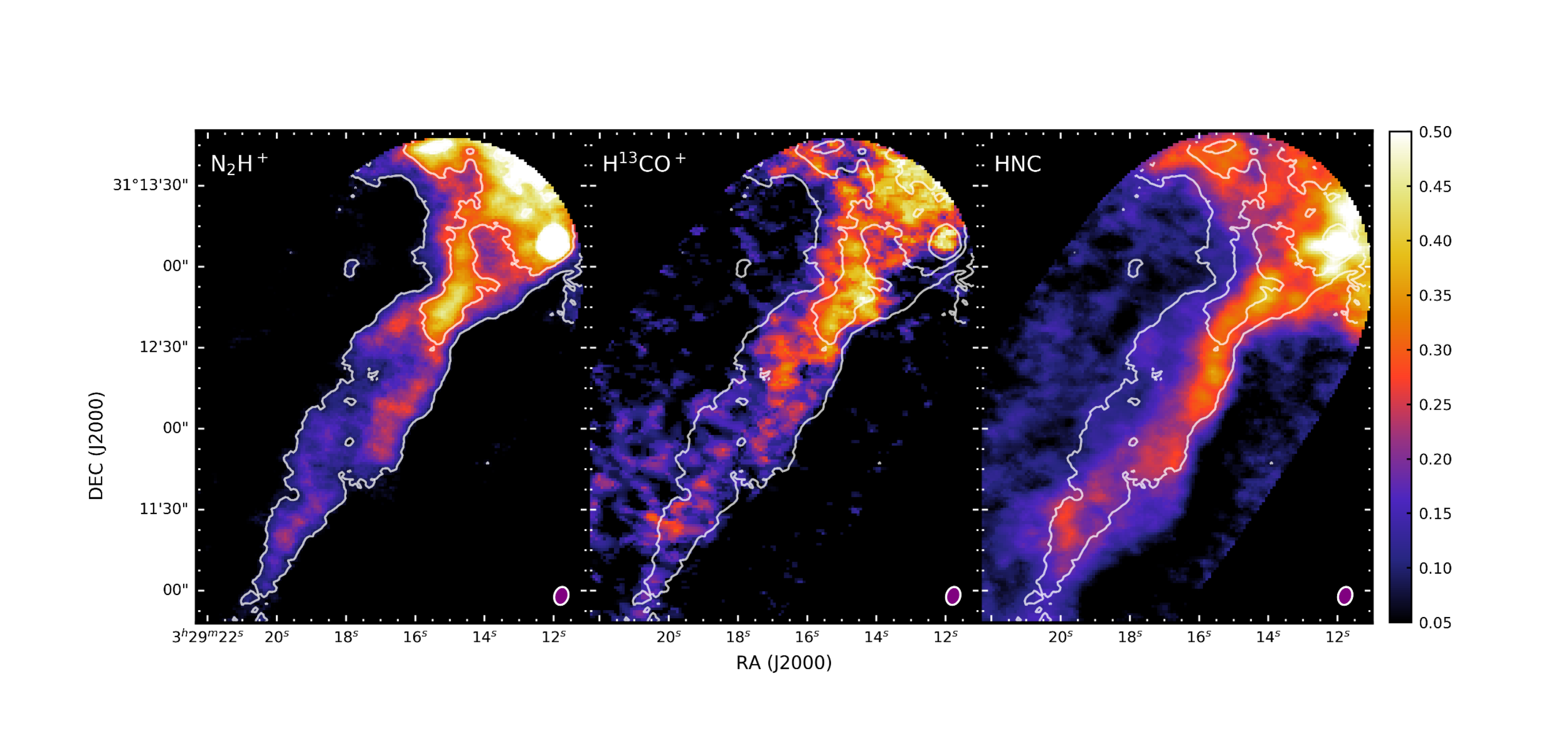}
\includegraphics[trim={1.5cm 1cm 2.2cm 1.75cm},clip,width=\textwidth]{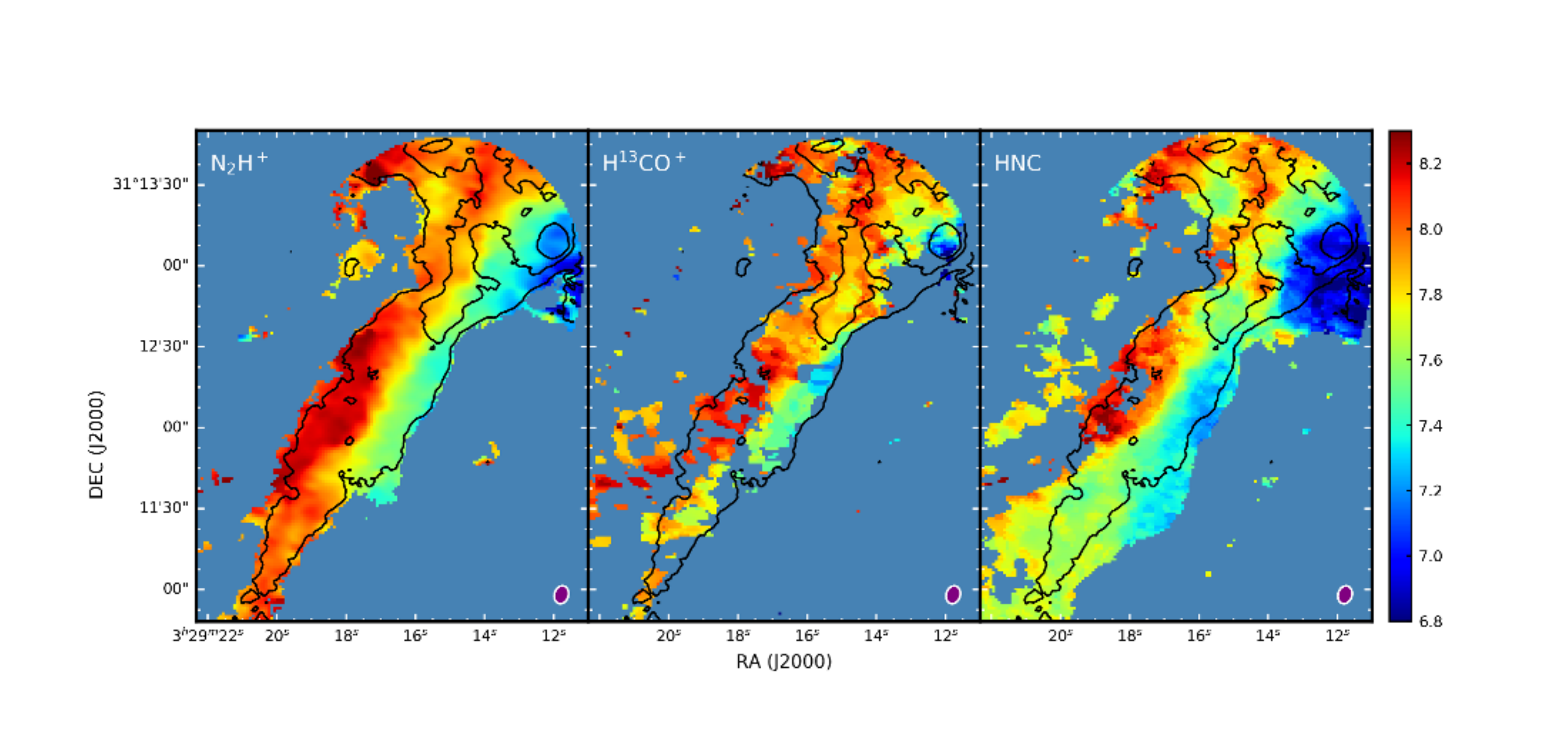}
\caption{Same as Fig. \ref{fig:SERPSNW} for NGC 1333 - SE region. The integrated intensity maps are in units of 6.0, 1.0 and 4.0  $Jy\ beam ^{-1}\ km\ s^{-1}$ for \nnh, \htco, and HNC respectively. The \nnh\ contours are at (1,3,5) $\times$ $\sigma$, where $\sigma$ = 0.4 $Jy\ beam ^{-1}\ km\ s^{-1}$.}
\label{fig:N1333SE}
\end{figure*}

\begin{figure*}
\centering
\includegraphics[trim={0cm 0cm 0cm 0.5cm},clip,width=0.9\textwidth]{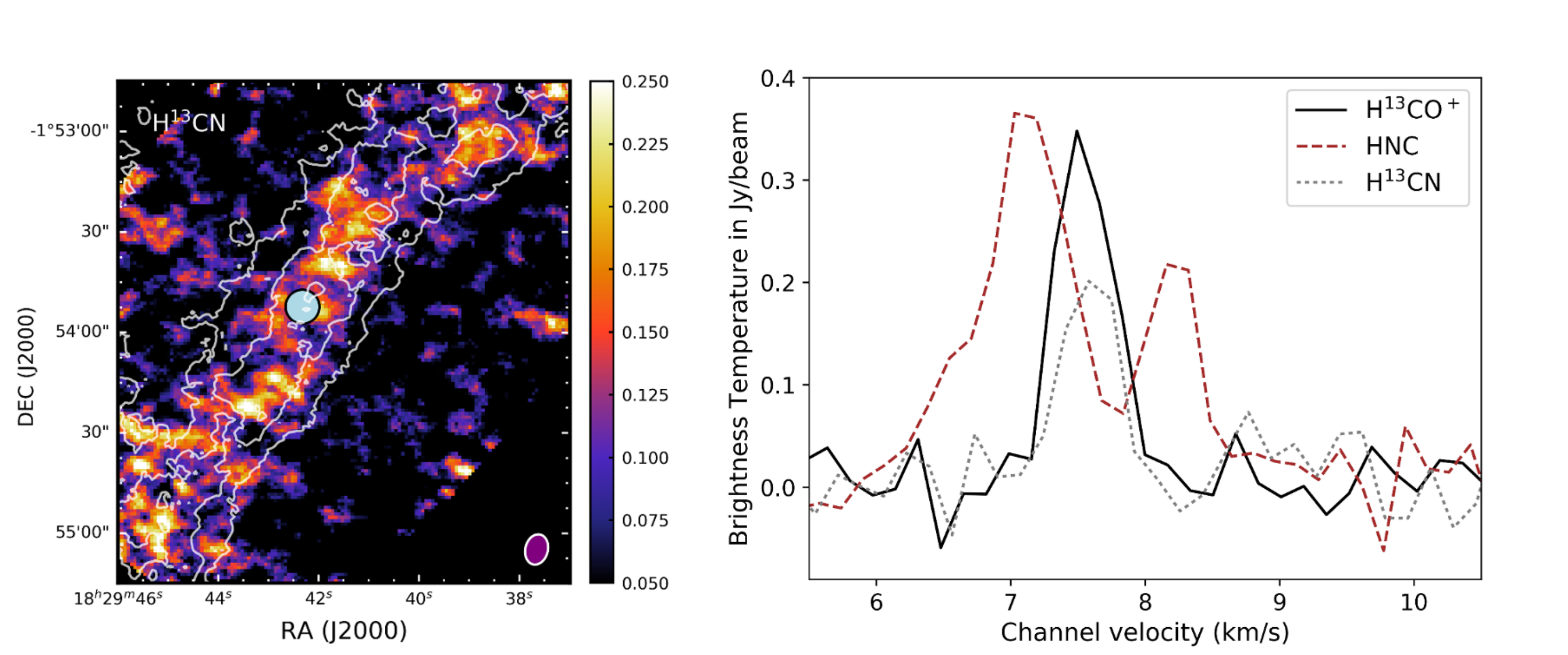}
\caption{\textit{Left}: Integrated intensity maps of Serpens South - NW filament using \htcn\ J=1-0 transition line. The map is in units $Jy\ beam ^{-1}\ km\ s^{-1}$. The \nnh\ integrated intensity contours are overlaid on the image as in Figure \ref{fig:SERPSNW}. The synthesized beam is shown at the bottom-right of the image. The blue circle marks the location where the spectra is taken for comparison with other molecules. \textit{Right}: \htco\ (black, solid), HNC (brown, dashed) and \htcn\ (grey, dotted) J=1-0 spectra taken at the location shown in the left figure. The peaks of \htco\ and \htcn\ coincide, but the HNC spectrum shows an absorption dip.}
\label{fig:SERPSNW2}
\end{figure*}

We display the moment maps for \htco\ and HNC and compare them with the CLASSy \nnh\ maps in Figures \ref{fig:SERPSNW}, \ref{fig:SERPSE}, \ref{fig:SERPME}, \ref{fig:SERPMS} and \ref{fig:N1333SE}. The \htcn\ maps are weak, having signal-to-noise ratio (SNR) of less than 4 along the filaments, which is not sufficient to do kinematic analysis. We present the \htcn\ integrated intensity map only for the Serpens South - NW region (Figure \ref{fig:SERPSNW2} \textit{left}).

We find that the \nnh\ emission corroborates well with the \htco\ emission, both for the integrated intensity and velocity maps. Although the \nnh\ emission intensity is more than 5 times brighter than the \htco\ emission, the RMS noise in the \nnh\ maps are also greater. So depending on the SNR of the \nnh\ maps in comparison to the \htco\ maps, in some cases we have more extended emission in \nnh\ and vice-versa. For example, in the Serpens South regions, the filaments are traced much more extensively in \nnh, while in the Serpens Main region, we detect more structure and trace the known structures over larger areas in \htco.

The HNC emission traces most of the structures seen in \nnh\ and \htco, but the relative emission varies in different regions of the maps. HNC emission also has greater SNR than the \htco\ maps by a factor of 2 to 4. In the Serpens South - NW region (Figure \ref{fig:SERPSNW}), it traces the NW part of the filament more than the central part, while in the Serpens Main - cloud center and E filament (Figure \ref{fig:SERPME}), the HNC emission is more extensive. In the NGC 1333 - SE region (Figure \ref{fig:N1333SE}), HNC traces the west side of the filament more than the east by a factor of two, although both in \htco\ and \nnh, the relative emission of the two sides are comparable.

The HNC velocity maps are noticeably different from the corresponding \htco\ and \nnh\ maps for most of the regions. This is because its spectrum shows absorption dips at many locations along the filaments. This results in two peaks on either side of the line center in \htco\ and \htcn\ emission; the relative strengths of the two HNC peaks depend on the velocity of the absorbing material in the line of sight (see Figure \ref{fig:SERPSNW2} \textit{right}). Since HNC is a main isotopic species with higher abundance, it is not surprising that it shows absorption in the J=1-0 transition as was previously found for HCN and \hco\ \citep{Storm2014, Lee2014}. However, there are some regions especially near the Serpens Main cloud core, where the HNC spectrum does not show absorption features. In these cases, there is a greater match in the moment maps between the different tracers. The variations in the HNC spectra are further discussed in Section \ref{sec:Abs}.

%% file: analysis.tex
\section{Analysis}
\label{sec:Analysis}

By themselves, the integrated intensity and velocity maps are insufficient to understand the finer structure in many of the studied regions. The \htco\ (J=1-0) channel maps are best suited for analysis especially in fields containing multiple sub-structures in close position or velocity proximity. This species is optically thin and has an isolated spectral line (corresponding to the J=1-0 transition), while at the same time it has sufficient SNR to detect the filaments. The complex hyperfine structures (in \nnh\ and HNC) or absorption features (in HNC) of multiple sub-structures can overlap, leading to degeneracy and challenges in disentangling the contributions of the separate components. However, these species have greater abundance than \htco\ and hence they can easily map extensive areas.

To disentangle the structures, we compare individual channel maps in \htco, and use position-velocity diagrams to check for emission continuity. In this section, we present \htco\ contour maps, averaged over 3-4 contiguous channels. These contours are overlaid on \textit{Herschel} column density maps obtained from the Gould Belt Survey Archive \citep{Kony2015, Pezzuto2012} to compare the individual structures with the dust emission. We also use the isolated hyperfine component of \nnh\ in regions having SNR $>$ 2 to confirm our identification of the velocity-coherent structures. We use other indicators such as variations in HNC absorption features and relative abundances to further support the identifications of structures in complex regions.

Following the identification of velocity-coherent sub-structures, we determine their extent using contours at half the peak emission for each channel. These contours over multiple channels are stacked, and we establish the lengths of the major and minor axes for the combined area within these stacked contours. The minor axis gives a representative width value, while the major axis gives a minimum length estimate (within the observed area). It is to be noted that this method only gives a rough estimate of the extent of the individual structures. We use a more formal method for determining the widths by Gaussian curve fitting of cuts in the integrated intensity maps (discussed in details in Section \ref{sec:Width}), but it is applicable only to filaments not having overlapping sub-structures. 

Only the structures having aspect ratios greater than 4 (as in \cite{Lee2014}) and at least 0.25 $pc$ long are considered in our velocity analyses. We identify 8 such `filamentary structures' over the 5 regions by this method (see Table \ref{tbl:Grad}). All of them are aligned with the filaments identified in \textit{Herschel} maps, although some are not resolved into multiple components in the low-resolution dust maps. Serpens South – NW is the least complex region having only one velocity coherent structure along the \textit{Herschel} filament. The Serpens South - E region, the Serpens Main - S region and the NGC 1333 – SE region each have two partially overlapping velocity coherent structures. Other authors have noted similar structures in observations and simulations and termed them as sub-filaments \citep{Smith2014} or as fibers \citep{Hacar2013}. We refer to them as sub-filaments or components and denote them by letters A and B. The Serpens Main - Cloud center and E filament region also has overlapping structures at the cloud center, but the Herschel filament east of the cloud does not have any identifiable sub-structure. 

Many of the filamentary structures show systematic variations in their velocity. In non-overlapping filamentary structures, the local velocity gradients are calculated directly from the Moment 1 maps. For the overlapping structures, we trace the \htco\ spectral peak over the extent of the filament, identifying the peak corresponding to each structure. Their velocity gradients are calculated locally using the \htco\ spectral peak velocity difference in different parts of the same velocity coherent structure. They are reported as gradients only if they satisfy the following conditions: (i) the velocity monotonically changes in one spatial direction (generally along the filament or perpendicular to it), (ii) the velocity difference is greater than or equal to 0.2 $km/s$ and (iii) the above conditions are satisfied for a distance of at least 4 beam-widths along the filament. From the multiple local velocity gradient values for each filamentary structure, we report the maximum, minimum and mean gradients (see Table \ref{tbl:Grad}). The error in the gradient calculation can be estimated from the velocity difference errors reported in the table. They are generally less significant than the variation in the gradient values over a filamentary structure.

For our analyses, we use distances of 436 $pc$ for the 4 Serpens regions \citep{OL2017}, and 300 $pc$ for the Perseus region \citep{PDist2014}. Both clouds have an average recessional velocity of $7.5-8$ $km/s$. Dust temperatures in these filaments range from 12 $K$ to 15 $K$ \citep{Rocca2015, Tanaka2013}. We use the \textit{Spitzer} catalog of YSOs \citep{Cat2015} and identified point sources in the \textit{Herschel} 70 $\micron$ maps corresponding to the studied regions to compare their locations with respect to the structures identified by us. 

We also calculate the mean \htco\ and \htcn\ velocity dispersions $\langle\sigma\rangle$ for the two isolated filaments, where we can determine the value unambiguously by taking statistics over a large part of the structure. The non-thermal velocity dispersion is calculated using the expression $\langle\sigma\rangle_{nt} = \sqrt{\langle\sigma\rangle^2 - kT/(\mu m_H)}$ where T is the kinetic temperature and $\mu m_H$ is the mass of each molecule of the tracer ($\mu$ = 30 for \htco\ and 28 for \htcn\, and $m_H$ is the the atomic Hydrogen mass). At typical filament temperatures of 12-15 $K$, the non-thermal dispersion is the main component of the overall dispersion. We compare this to sound speeds $c_s$ $\approx$ 0.22-0.25 $km/s$ at these temperatures.

\begin{table*}
\centering
\input{Tbl_Gradient.tex}
\caption{Widths and gradient statistics of filamentary structures using \htco\ maps}
\label{tbl:Grad}
\end{table*}

\subsection{Serpens South - NW Region}

\begin{figure*}
\centering
\includegraphics[width=0.4\textwidth]{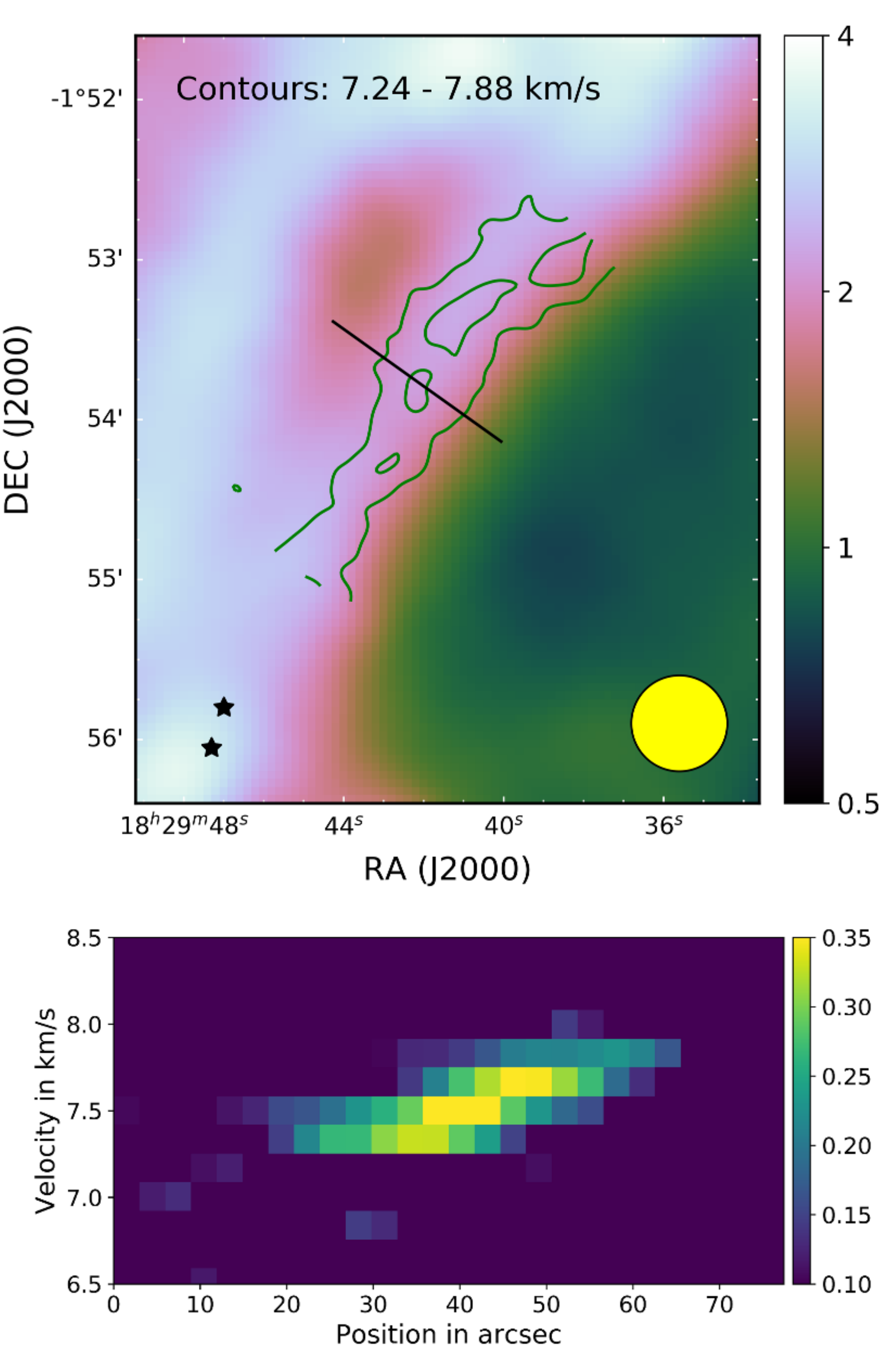}
\caption{\textit{Top}: \textit{Herschel} column density map ($10^{22}$ $cm^{-2}$) with \htco\ contours overlaid on it for the Serpens South - NW region. The contours are at (2,4) $\times$ $\sigma$ ($\sigma$ = 0.06 $Jy/beam$) for \htco\ emission averaged over 4 channels between 7.24 $km/s$ and 7.88 $km/s$. The black line corresponds to the cut for which the position-velocity diagram is presented in the lower panel. The black stars denote the locations of known Class 0/I YSOs. The yellow circle at the bottom-right represents the resolution of the column density map, corresponding to the \textit{Herschel} 500 \micron\ beam. \textit{Bottom}: Position-velocity plot for a cut across the filament,showing a single velocity-coherent component. The slope indicates the velocity gradient.}
\label{fig:AnaSERPSNW}
\end{figure*}

Figure \ref{fig:AnaSERPSNW} (top) shows the \htco\ contours of the Serpens South - NW region, averaged over 4 channels. This region has no evident sub-structures, as can be confirmed from the position-velocity diagram for a representative cut across the filament. There is a velocity gradient of about 0.24 $km/s$ over the 0.04 $pc$ width of the filament both in the \htco\ and \nnh\ emission. This is seen all along the 0.4 $pc$ length of the filament in the mapped area  (Figure \ref{fig:SERPSNW}). The mean \htco\ velocity dispersion for this filament (0.21 $km/s$) is transonic. There are no YSOs identified along the filament, although there are multiple Class 0/I sources near the filamentary hub at the south-eastern end of the studied region. 

\subsection{Serpens South - E Region}

\begin{figure*}
\centering
\includegraphics[width=0.7\textwidth]{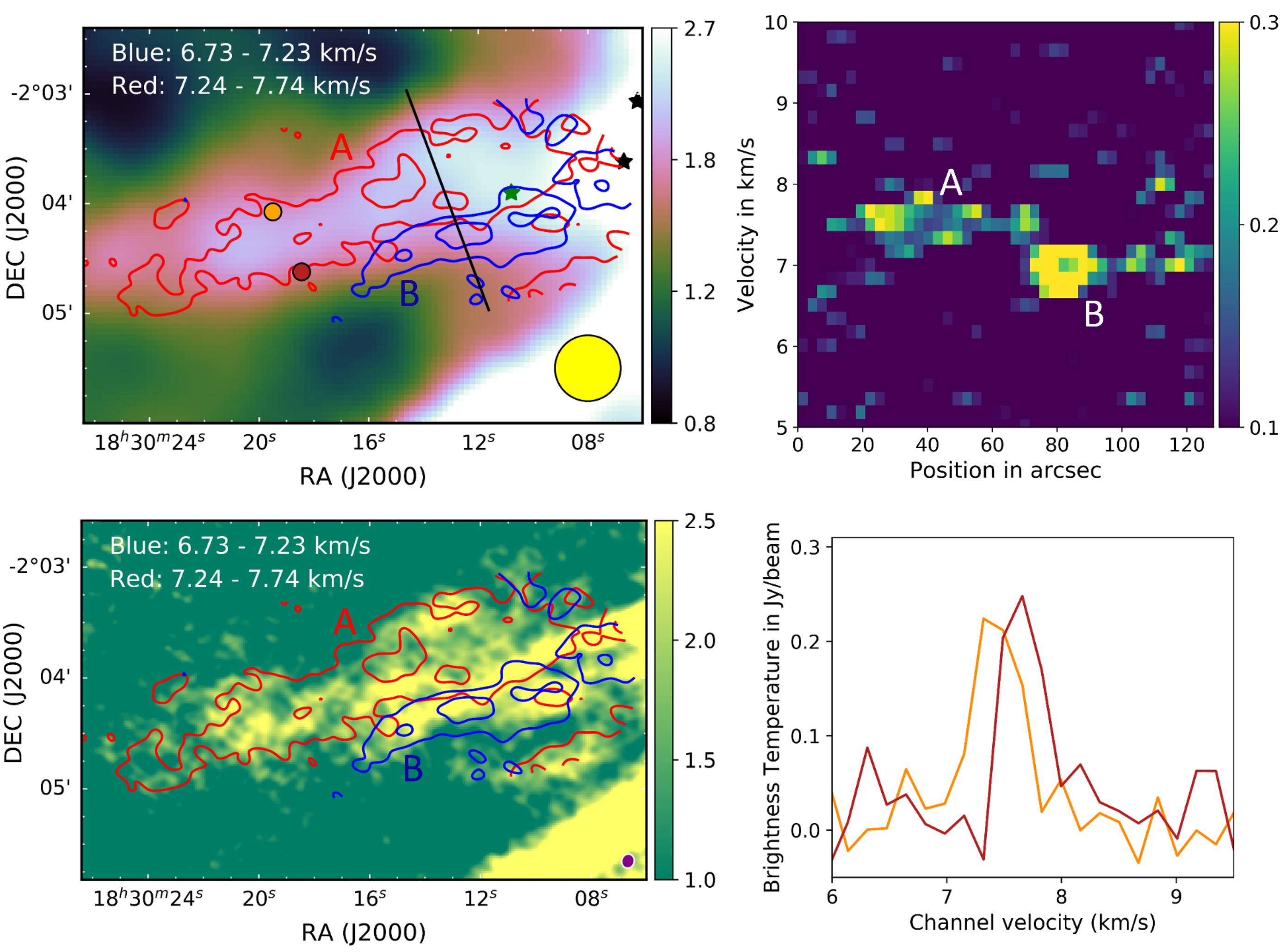}
\caption{\textit{Upper-left}: \textit{Herschel} column density map ($10^{22}$ $cm^{-2}$) with \htco\ contours overlaid on it for the Serpens South - E region. The red contours correspond to emission averaged over 3 channels centered at 7.49 $km/s$, while the blue contours correspond to emission averaged over 3 channels centered at 6.98 $km/s$. The red and blue contours represent the two sub-filaments `A' and `B'. The contours are at (2,4,6) $\times$ $\sigma$ ($\sigma$ = 0.08 $Jy/beam$). The black line shows the cut for which the position-velocity diagram is presented in the upper-right panel. The dark red and orange circles are two regions where the \htco\ spectrum is taken and presented in the lower-right panel. The black and green stars denote the locations of known Class 0/I and Flat spectrum YSOs respectively. The yellow circle represents the resolution of the \textit{Herschel} map. \textit{Upper-right}: Position-velocity plot for a cut across the filament showing the two velocity-coherent components. `A' is wider and fainter than `B'. \textit{Lower-left}: \nnh\ integrated intensity map ($Jy\ beam ^{-1}\ km\ s^{-1}$) with the \htco\ contours corresponding to the two sub-structures overlaid on it. The synthesized beam for the \nnh\ map is shown at the bottom-right. \textit{Lower-right}: \htco\ spectrum at two locations marked by circles of the same color in the left-most panel. This shows the velocity gradient which is present in the eastern half of the filament, only in component `A'.}
\label{fig:AnaSERPSE}
\end{figure*}

Figure \ref{fig:AnaSERPSE} shows two sub-filaments in the Serpens South - E region running in the east-west direction - a weaker sub-filament (A; red contours) which is wider and a parallel narrow sub-filament (B; blue contours) having stronger emission. These are easily separable in the position-velocity diagram (middle panel of Figure \ref{fig:AnaSERPSE}). There is partial line-of-sight overlap between these two components. There is a large line-of-sight velocity difference over the region. It varies between 6.6 $km/s$ and 8.0 $km/s$ according to the \htco\ and \nnh\ maps (see Figure \ref{fig:SERPSE}). Sub-filament `A' is more red-shifted and has a velocity gradient across it, while `B' is blue-shifted and has no consistent gradient across it. In the \htco\ maps, because of low SNR, the gradient could be determined only in the east part of `A'. The right panel of Figure \ref{fig:AnaSERPSE} shows spectra at two representative locations across the filament. The gradient is about 4.2 $km\ s^{-1} pc^{-1}$.

The HNC emission has absorption features in sub-filament `A' , but absorption is absent or minimal in `B'. The sharp velocity and intensity gradients near the brightest emission combined with the differences in the HNC absorption across the two regions are also in support of the multiple structure interpretation. Two Class 0/I/Flat YSOs are identified in this region closer to the cloud center. Both are in the region where the two sub-filaments overlap in the line-of-sight, and thus cannot be associated with one or the other.

Using the \nnh\ maps, \cite{FL2014} could not distinguish between the two velocity coherent components and concluded a large velocity gradient across a single filament structure. Even though the Moment 0 and velocity maps of \nnh\ are similar to \htco\, with the \nnh\ maps having a higher SNR than \htco\ (for this region), the hyperfine structure of \nnh\ prevents analysis using position-velocity diagrams to distinguish the velocity coherent structures. The gradient in the eastern part although visible in the \nnh\ maps as well, has a lesser magnitude and opposite sense compared to the inaccurately identified gradient in the central and western part, and was thus not reported in \cite{FL2014}. With the knowledge of the two-component structure using the \htco, we find that the isolated hyperfine line of \nnh\ (J=1-0, F=0-1) shows similar trends as \htco, but by itself this line has much lesser SNR than \htco, and is too weak to map the kinematics of the entire region. 

\subsection{Serpens Main - Cloud Center and E Filament}

\begin{figure*}
\includegraphics[width=\textwidth]{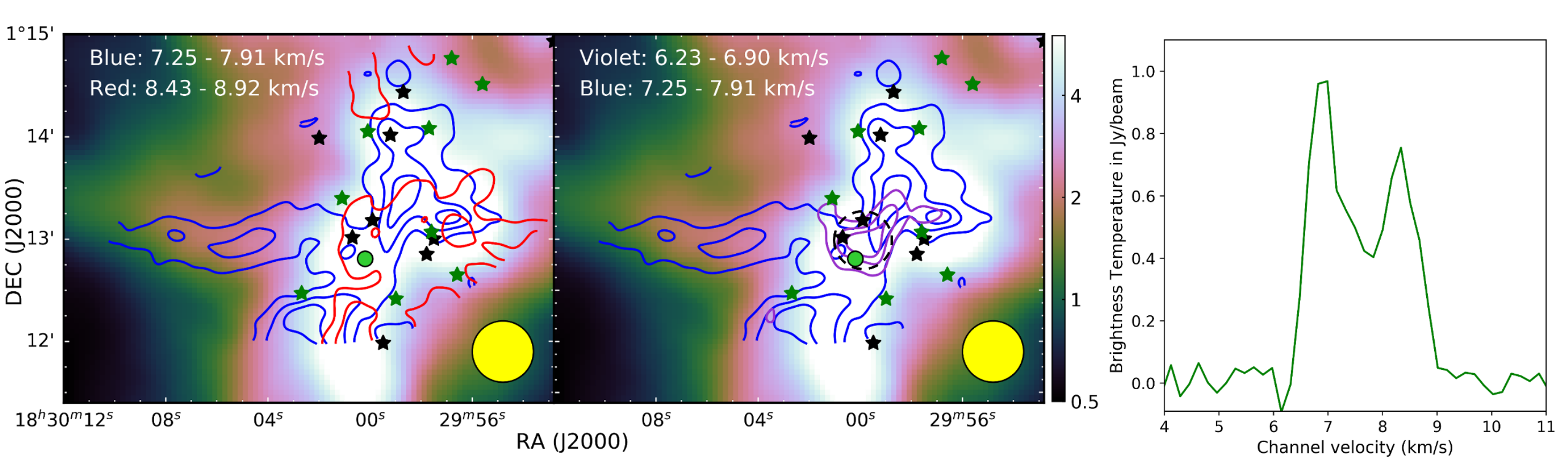}
\caption{\textit{Left}: \textit{Herschel} column density map ($10^{22}$ $cm^{-2}$) for Serpens Main - cloud center and E filament, with \htco\ contours overlaid on it for emission averaged over channels centered at 8.68 $km/s$ (red) and at 7.58 $km/s$  (blue). The contours are at (2,4,6) $\times$ $\sigma$ ($\sigma$ = 0.1 $Jy/beam$). The two sets of contours correspond to the two sets of structures identified in the region. The green circle near the center marks the location where the \htco\ spectrum is taken and presented in the right-most panel. The black and green stars denote the locations of known Class 0/I and Flat spectrum YSOs respectively. The yellow circle represents the resolution of the \textit{Herschel} map. \textit{Middle}: Same as the left panel for \htco\ channels centered at 7.58 $km/s$ (blue contours) and 6.57 $km/s$ (violet contours). In spite of the large velocity spread, these two components are part of the same structure. The two sets of contours however indicate the flows towards a potential well in this region at the center of the violet contours. The dashed black circle marks the intersection region of the flows. \textit{Right}: \htco\ spectrum at the locations marked by the green circle in the left and middle panels. This shows that there are two sets of velocity-coherent structures separated by about 1.4 $km/s$.}
\label{fig:AnaSERPME}
\end{figure*}

This is a complex region mapping the south-east part of the Serpens Main hub, with emission extending in various directions. On comparing to the larger scale structures in the region using \textit{Herschel} maps, we see that only two parts of the emission form long filaments, one in the east, and the other in the south. The east filament is mapped completely in this region. The filament in the south is completely mapped in the Serpens Main - S region (discussed in the following section). 

In the cloud center region, there is a large range of line-of-sight velocities from 6 $km/s$ to 9 $km/s$. Using \htco\ channel maps, two sets of structures associated with different velocities are identified within the hub (Figure \ref{fig:AnaSERPME} \textit{right}). Of these, the velocities above 8 $km/s$ correspond to a separate velocity-coherent structure (marked in red in the left panel of Figure \ref{fig:AnaSERPME}) while another set of structures have velocities in the range of 6 to 8 $km/s$. This second set includes a convergent point at ($\alpha, \delta$) = (18:29:59.8, +1:13:00) where three flows seem to intersect from east, south-east and north-west respectively (middle panel of Figure \ref{fig:AnaSERPME}). We call them flows because they have a velocity gradient, such that their ends closer to the hub is highly blue-shifted, possibly indicating acceleration in a potential well. The HNC spectrum has no absorption dips in the blue-shifted set of structures, but has very strong dips in the red-shifted regions.

Of these flows, the one from the east is identified as a filament in our analysis. Most of the filament is isolated from the other structures and so can be analyzed from the moment maps. The filament has a velocity gradient of 1.1 $km/s$ over its length of 0.23 $pc$. The gradient vector is oriented along the filament close to the hub (10.3 $km\ s^{-1} pc^{-1}$), but its magnitude reduces by a factor of four 0.2 $pc$ away from the cloud center. Moving further from the hub, the gradient direction rotates by about 90 degrees such that it is oriented almost across the filament near the easternmost extreme. The mean velocity dispersion over this filament is 0.33 $km/s$, which is marginally supersonic.

In the central hub, three point sources SMM1, SMM3 and SMM4 are identified in the continuum map, but they are not closely associated with the multi-filament intersection point. However, on comparing with the \textit{Spitzer} YSO catalog \citep{Cat2015}, two Class I/0 YSOs are identified within 5000 $AU$ projected distance from this point. Additionally, an outflow identified in previous CO maps \citep{Davis1999} and the CLASSy \hco\ map can be traced to be originating close to the filament intersecting region. 

\subsection{Serpens Main - S Region}

\begin{figure*}
\centering
\includegraphics[width=1.0\textwidth]{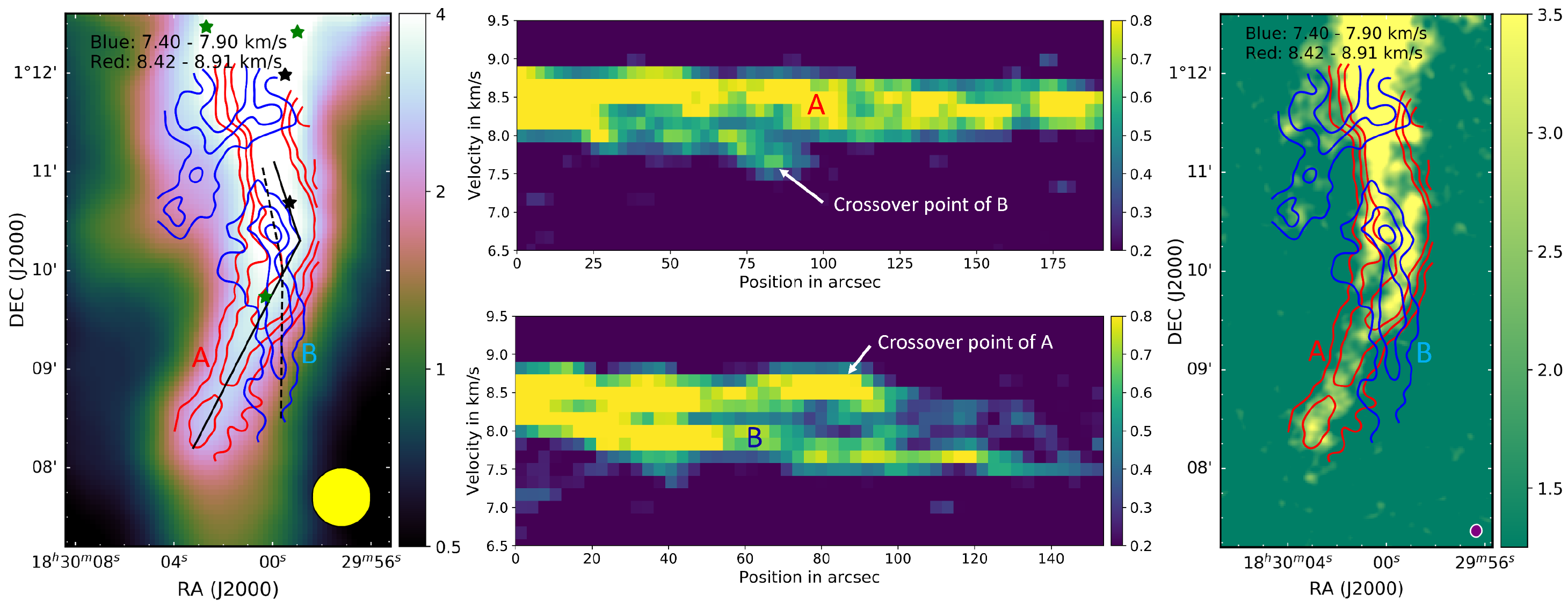}
\caption{\textit{Left}: \textit{Herschel} column density map ($10^{22}$ $cm^{-2}$) with \htco\ contours overlaid on it for the Serpens Main - S region. The red and blue contours correspond to emission averaged over 3 channels centered at 8.66 $km/s$ and 7.65 $km/s$ respectively. They represent the two sub-filaments `A' and `B'. The contours are at (2,4,6) $\times$ $\sigma$ ($\sigma$ = 0.1 $Jy/beam$).  The channels in between 7.90 $km/s$ and 8.42 $km/s$ have emission corresponding to both the sub-filaments and are inseparable in the channel maps. The black solid and dashed lines correspond to the cuts along the lengths of `A' and `B', for which the position-velocity diagrams are presented in the middle panels. The black and green stars denote the locations of known Class 0/I and Flat spectrum YSOs respectively. The yellow circle represents the resolution of the \textit{Herschel} map. \textit{Middle-top}: Position-velocity plot for a cut along `A' (solid line in left panel), showing its velocity range from 8.2 to 8.7 $km/s$ with no evident gradients. The left half of the image also shows the emission from sub-filament `B' before it crosses over and the separation of the components increase in the projected sky plane. \textit{Middle-bottom}: Position-velocity plot for a cut along `B' (dashed line in left panel), showing its velocity changing from 7.5 to 8.5 $km/s$. `B' is weaker than `A' and so there is substantial contribution from `A' along the cut. \textit{Right}: \nnh\ integrated intensity map ($Jy\ beam ^{-1}\ km\ s^{-1}$) with the \htco\ contours corresponding to the two sub-structures overlaid on it. The synthesized beam for the \nnh\ map is shown at the bottom-right. }
\label{fig:AnaSERPMS}
\end{figure*}

The northern part of this region maps into part of the cloud center of Serpens Main. In this upper part, two velocity components are identified at about 7.3 $km/s$ and 8.4 $km/s$ respectively. The relatively blue-shifted component has a greater spatial extent and does not show any absorption in the HNC spectrum compared to the relatively red-shifted component. 

This region also has two sub-structures along the filament, which originates from the hub in the north (see Figure \ref{fig:AnaSERPMS}) and are more than 0.3 $pc$ in length. These velocity-coherent sub-filaments `A' and `B' are parallel to each other and cross-over in the projected sky plane at ($\alpha, \delta$) = (18:29:59.5, +1:09:45), corresponding to an emission peak in the integrated intensity maps (see Figure \ref{fig:SERPMS}). The position-velocity diagrams for cuts along the two sub-filaments reveal their velocity distribution. Sub-filament `A' has velocities in the range $8.2-8.7$ $km/s$ with no evident gradients. Sub-filament `B' has a velocity gradient along the filament such that the line-of-sight velocities are 7.5 $km/s$ at the south end of the filament, and 8.5 $km/s$ close to the cloud center. As in the Serpens Main - E filament (discussed in the previous section), this sub-filament also indicates accelerated flow closer to the cloud center. The gradient changes from 1.2 $km\ s^{-1} pc^{-1}$ to 5.0 $km\ s^{-1} pc^{-1}$, going towards the cloud center. 

The \nnh\ emission identifies both these sub-filaments, but does not trace the lower half of `B' because of insufficient SNR. In their kinematic analysis, \cite{Lee2014} used a single velocity component fit for most regions. They used a two velocity component fit only in regions where there is a large velocity difference between the components ($\geqslant$ 1 $km/s$) that could be resolved unambiguously. They detected two sub-filaments using the integrated intensity maps since the sub-filaments had well-defined ridges, spatially resolved by the CARMA beam. However they used a single component fit to obtain the velocity maps because the velocity difference between the components in the northern part is $\leqslant$ 0.5 $km/s$, and were thus unable to capture the extent of the two components in the overlapping regions. 

The velocity gradient observed across the filament in the lower half of the \htco\ first moment map (see Figure \ref{fig:SERPMS}) is not a true gradient; it is an effect of the overlapping sub-filaments. This case is similar to that observed by \cite{Beu2015} in the dense filament IRDC 18223, and by \cite{Moeckel2015} in simulations.

South of the intersection points of `A' and `B', the spectrum corresponding to emission from `A' is wide, and at places divides into two peaks separated by 2-3 channels. In the HNC integrated intensity map, the width of `A' is lesser than in \htco, strengthening the case for an additional sub-structure in this region possibly having different physical parameters. However, there is insufficient evidence in our data to identify this sub-structure with certainty. Three Class 0/I/Flat YSOs are identified along the length of the filament. Of them, one Flat spectrum YSO is located close to the cross-over point of the sub-filaments.

\subsection{NGC 1333 - SE Region}

\begin{figure*}
\centering
\includegraphics[width=0.8\textwidth]{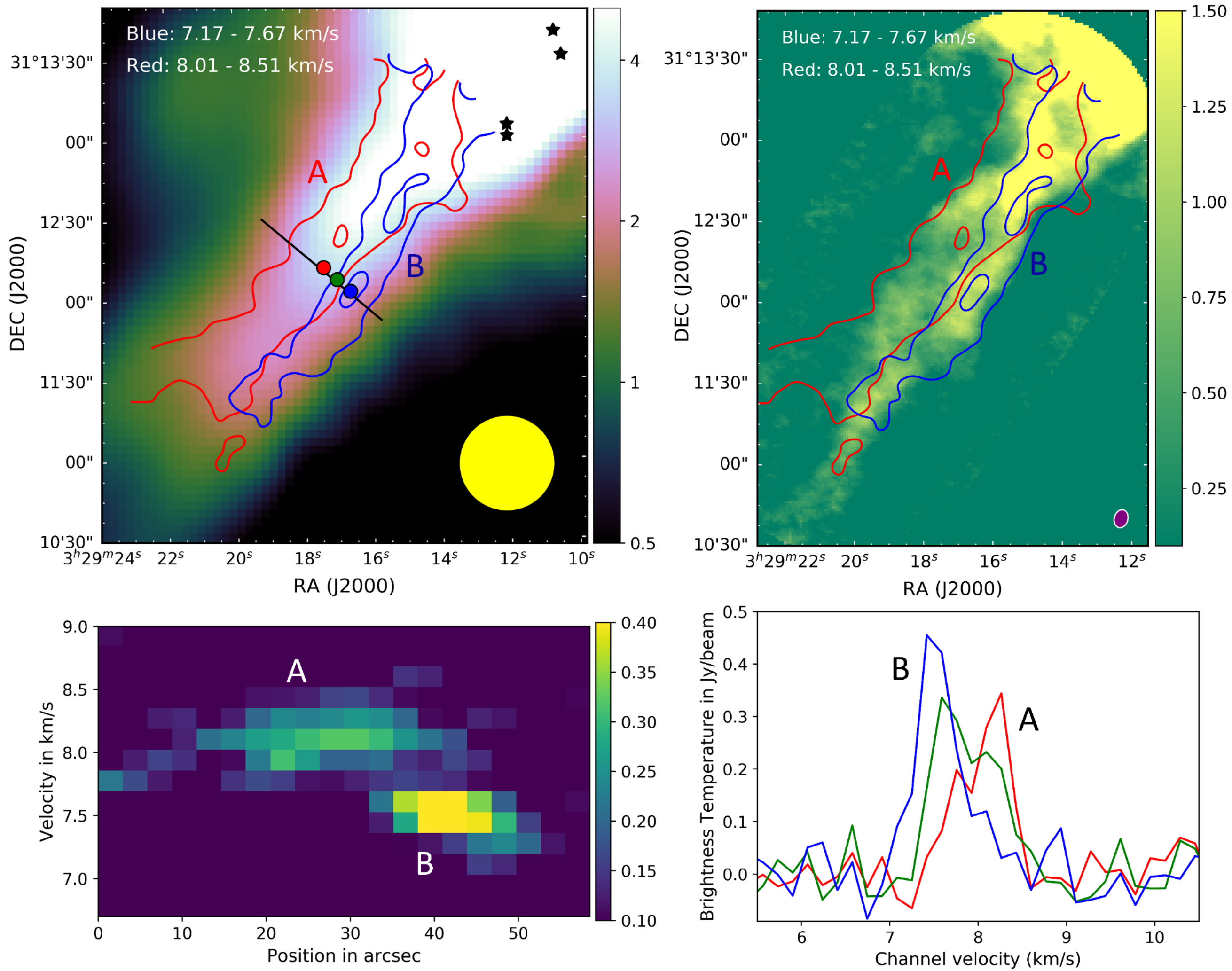}
\caption{\textit{Upper-left}: \textit{Herschel} column density map ($10^{22}$ $cm^{-2}$) with \htco\ contours overlaid on it for the NGC 1333 - SE region. The red and blue contours  correspond to emission averaged over 3 channels centered at 8.26 $km/s$ and 7.42 $km/s$ respectively. The red and blue contours represent the two sub-filaments `A' and `B', but not their true extent because the channels in between have emission from both the components that are inseparable in the channel maps. The contours are at (2,4,6) $\times$ $\sigma$ ($\sigma$ = 0.075 $Jy/beam$). The black line shows the cut for which the position-velocity diagram is presented in the lower-left panel. The red, green and blue circles are regions where the \htco\ spectrum is taken and presented in the lower-right panel. The black stars denote the locations of known Class 0/I YSOs. The yellow circle represents the resolution of the \textit{Herschel} map. \textit{Upper-right}: \nnh\ integrated intensity map ($Jy\ beam ^{-1}\ km\ s^{-1}$) with the \htco\ contours corresponding to the two sub-structures overlaid on it. The synthesized beam for the \nnh\ map is shown at the bottom-right. \textit{Lower-left}: Position-velocity plot for a cut across the filament showing the two overlapping velocity-coherent components. `A' has a fairly constant velocity, while `B' has a velocity gradient across it. \textit{Lower-right}: \htco\ spectrum at three locations marked by circles of the same color in the left panel. The spectra indicate that there are two velocity components, the relative intensities of which change as we move across the filament. This is also associated with a shift in the spectral peak of `B' by 0.33 $km/s$. This is how we distinguish between the emission peaks corresponding to the two components and appropriately determine their independent velocity gradients.}
\label{fig:AnaN1333SE}
\end{figure*}

This region has a filament with two sub-structures running parallel to each other from north-west to south-east. As shown in Figure \ref{fig:AnaN1333SE}, the eastern sub-filament (A) has a velocity in the range $8.1-8.4$ $km/s$ with a fork towards the south. The western sub-filament (B) has a velocity gradient across its 0.03 $pc$ width changing gradually from about 7.8 $km/s$ to 7.5 $km/s$. As explained at the beginning of this section, we measure the gradient using the spectral peak locations corresponding to the same velocity-coherent structure. This is shown in the bottom-right panel of Figure \ref{fig:AnaN1333SE}. In some sections of sub-filaments `A', there is a small gradient in the opposite sense compared to that for `B'. 

The relative intensities of the two sub-filaments are comparable, both with \nnh\ and \htco. However the HNC emission from the right sub-filament is about 2 times brighter than the left. The difference in the HNC emission compared to the other two tracers strengthens the interpretation that there are two distinct sub-filaments having different physical parameters.

The four YSOs corresponding to the components of IRAS 4 are located in the top-right part of the mapped region. They appear bright in the CLASSy-II continuum (360 $Jy/beam$ and 130 $Jy/beam$ peak intensities respectively) maps. Neither of the  outflow axes from these protostars \citep{Koumpia2016} are oriented along the filament in this region. \cite{Stephens2017} postulated that the lack of correlation between outflow axis directions and the filament orientations indicates that the angular momentum axis of a protostar may be independent of the large-scale structure.

%% file: Tbl_Gradient.tex
\newcolumntype{P}[1]{>{\centering\arraybackslash}p{#1}}

\begin{tabular}{ p{10em} P{7em} p{4em} P{5em}  P{6.5em} P{2em} P{2em} P{2em}} 
\toprule
\multirow{2}{10em}{Filament (sub-filament)} & \multirow{2}{7em}{\centering{Representative width$^a$ ($pc$)}} & \multirow{2}{4em}{Gradient Direction$^b$}& \multirow{2}{5em}{\centering{$\langle \Delta v \rangle $ ($km/s$)}} & \multirow{2}{6.5em}{\centering{Distance along filament$^c$ ($pc$)}} & \multicolumn{3}{c}{$\nabla v$ ($km\ s^{-1} pc^{-1}$)} \\ \cline{6-8}
& & & & & Min & Max & Mean \\
\midrule 

Serpens South - NW & 0.06 & across & 0.24 $\pm$ 0.07 & 0.19 & 3.9 & 6.8 & 5.5\\
Serpens South - E (A) & 0.08 & across & 0.22 $\pm$ 0.08 & 0.11 & 3.6 & 5.3 & 4.2\\
Serpens South - E (B) & 0.04 & - & & & & & \\
\multirow{2}{10em}{Serpens Main - E} & \multirow{2}{7em}{\centering 0.04} & across & 0.54 $\pm$ 0.09 & 0.07 & 5.7 & 12.0 & 8.9\\
& & along & 1.06 $\pm$ 0.06 & 0.23 & 2.7 & 10.3 & 4.6 \\
Serpens Main - S (A) & 0.07 & - & & & & & \\
Serpens Main - S (B) & 0.05 & along & 0.78 $\pm$ 0.15 & 0.29 & 1.2 & 5.0 & 2.6\\
NGC 1333 - SE (A) & 0.05 & - & & & & &\\
NGC 1333 - SE (B) & 0.03 & across & 0.30 $\pm$ 0.14 & 0.09 & 8.1 & 12.2 & 9.9 \\
\bottomrule
\end{tabular}

\raggedright{\footnotesize{$^a$ The widths are calculated from the extent of contours in channel maps at half the peak emission. All the values vary in a range of about 0.03 $pc$ over the length of the filament. \newline $^b$ Only monotonically changing velocities $\Delta v$ $\geqslant$ 0.2 $km/s$ over a part of the filament $\geqslant$ 4 beam widths long are considered as gradients. \newline $^c$ Distance along filament spine over which gradient statistics are taken, both for perpendicular and parallel gradients.}}

%% file: discussion.tex
\section{Discussion}
\label{sec:Disc}

On comparing the CLASSy-II observations with the \textit{Herschel} dust continuum maps and the CLASSy \nnh\ maps, we find that the structure traced by all the maps are similar on the large scale. However on smaller scales ($<$ 0.1 $pc$), the finer structure of these filaments becomes evident in the CARMA maps. In many of the regions we identify multiple sub-structures, instead of a single uniform filament identified in the \textit{Herschel} maps. 

Using just the CLASSy \nnh\ maps, it was argued that the finer structure could be real or could alternatively be due to \nnh\ abundance variations caused as a result of \nnh\ depletion by CO in less dense regions \citep{Bergin2001}. However, the similarity between the structures traced by \nnh\ and \htco\ at the CLASSy resolution scale imply that the morphology and kinematics determined from these maps truly represent the dense gas distribution and are unlikely to be arising from chemical selectivity. There are some differences in the relative intensities of the structures traced by HNC, which could be an effect of relative abundance, temperature, density or a combination of them.

\subsection{Morphology}

The different regions studied in this paper indicate that despite the variety of filament structure, there are many common features. In this sub-section, we discuss the different aspects of filament morphologies and their implications. In some cases, rigorous analysis is only possible for the well isolated filaments, i.e. for Serpens South - NW filament and Serpens Main - E filament.

\subsubsection{Physical parameters of tracers}

Single transitions can be used to determine physical parameters like column density in molecular clouds only if we assume thermalization. Thus if we assume that a single excitation temperature $T_{ex}$ defines the level populations of a molecule, we can use the integrated intensity to calculate the total column density in the optically thin limit using the formula \citep{Goldsmith1999} 
\begin{equation}
\label{eqn:1}
N^{thin} = N^{thin}_u \frac{Q}{g_u e^{-E_u/kT_{ex}}} = \frac{8\pi k\nu^2}{hc^3 A_{ul}} \frac{\sum{g_i e^{-E_i/kT_{ex}}}}{g_u e^{-E_u/kT_{ex}}} \int_{-\infty}^{\infty} T_b dv
\end{equation}
where $c$, $k$ and $h$ are the speed of light constant, the Boltzmann constant and the Planck constant respectively. The transition frequency is $\nu$ and $A_{ul}$ is the Einstein A coefficient corresponding to the transition. Q is the partition function, which is assumed to be a function of a single variable $T_{ex}$. $g_i$ and $E_i$ are respectively the degeneracy and energy of the $i$th energy level. The subscripts $u$ and $l$ represent the upper and lower levels of the transition respectively. The integral represents the integrated line intensity with $T_b$ as the observed brightness temperature in $K$ and $dv$ as the channel width in $km/s$. Here we assume a unity beam filling factor. Further, a correction factor of $\tau/(1-e^{-\tau})$ is multiplied if the transition is optically thick. This opacity $\tau$ can be determined using the radiative transfer equation
\begin{equation}
\label{eqn:2}
T_b = \frac{h\nu}{k}\Big(\frac{1}{e^{h\nu /kT_{ex}} - 1} - \frac{1}{e^{h\nu /kT_{bg}} - 1}\Big)[1-e^{-\tau}]
\end{equation}
where $T_{bg}$ is the background radiation 2.73 $K$.

Because of the limitations of the analysis and the presence of overlapping structures in many regions further complicating the analysis, we present the results only for the Serpens South - NW filament ridge in \htco\ and \nnh. The lower limit of $T_{ex}$ for \nnh\ is estimated from the observed brightness temperature ($\sim$ 6 $K$). This $T_{ex}$ limit is also applicable to \htco\ since it has a critical density similar to that of \nnh. For the upper limit of $T_{ex}$, we use the maximum kinetic temperature from the dust temperature maps. Based on this, we use representative $T_{ex}$ values in the range 6-15 K. Using these equations and the assumed range of $T_{ex}$ values, we obtained column densities of $0.9-1.1$ $\times$ $10^{12}$ $cm^{-2}$ for \htco\ and $7.9-10.3$ $\times$ $10^{12}$ $cm^{-2}$ for \nnh\ along the ridge of the isolated filament in Serpens South - NW region. The column density values can be averaged over areas equal to the \textit{Herschel} beam size and compared to the H$_2$ column densities of $\sim$ 2.0 $\times$ $10^{22}$ $cm^{-2}$ obtained from \textit{Herschel} maps to get molecular abundances. By this method, we calculate abundances of $2.7-4.6$ $\times$ $10^{-11}$ for \htco, and $2.2-3.3$ $\times$ $10^{-10}$ for \nnh. 

We can also use the `large velocity gradient' (LVG) approximation to get a lower bound on the density of the region. Using the \htco\ observed brightness temperature and a maximum value of $\tau$ = 0.8 (from Equation \ref{eqn:2}) corresponding to $T_{ex}$ = 6 $K$, we get a minimum gas density of 3.0 $\times$ $10^5$ $cm^{-3}$ along the Serpens South - NW filament ridge. From the LVG model, we also obtain a minimum \htco\ column density of 0.9 $\times$ $10^{12}$ $cm^{-2}$ corresponding to the mean brightness temperature of 1.9 $K$. This estimate matches well with the analytically obtained column density in the previous paragraph.

\subsubsection{Filament widths in comparison with \textit{Herschel} dust maps}
\label{sec:Width}

Publications from the Gould Belt survey argued that all filaments have a similar width of about 0.1 $pc$ \citep{Arzou2011, Andre2010, Palm2013} with a narrow distribution. Other authors like \cite{Juvela2012} and \cite{Henn2012} have however reported larger widths of about 0.3 $pc$, while \cite{Pan2016} reported a lesser width of 0.06 $\pm$ 0.04 $pc$. \cite{Ysard2013} reported widths varying by a factor of 4, while \cite{Pan2017} concluded that a single characteristic width of filaments is inconsistent with observations, and that the narrow distribution is an averaging effect.

Earlier in Section \ref{sec:Analysis}, we reported the widths for the individual filamentary structures. Here we use a more systematic approach to calculate the widths of filaments using \htco\ and dust column density maps and compare between them. Since \htco\ is optically thin, the integrated intensity map scales directly with the column density (Equation \ref{eqn:1}). Hydrogen column density maps were obtained from the \textit{Herschel} Gould Belt archive. We take parallel cuts across the filaments at multiple points along its length, about 7\arcsec\ apart (comparable to beam size), average over the cuts and fit a Gaussian to get the full-width at half-maximum (FWHM). The deconvolved width, $W_d$, is obtained using the expression $W_d = \sqrt{(FWHM)^2-(HPBW)^2}$ where HPBW is the half-power beam width for the maps \citep{Kony2015}. We use distance estimates mentioned in Section \ref{sec:Analysis} to obtain the filament widths in parsecs. We also fit a Gaussian to each individual cut in order to obtain a range for the filament widths.

\begin{figure*}
\centering
\includegraphics[width=0.9\textwidth]{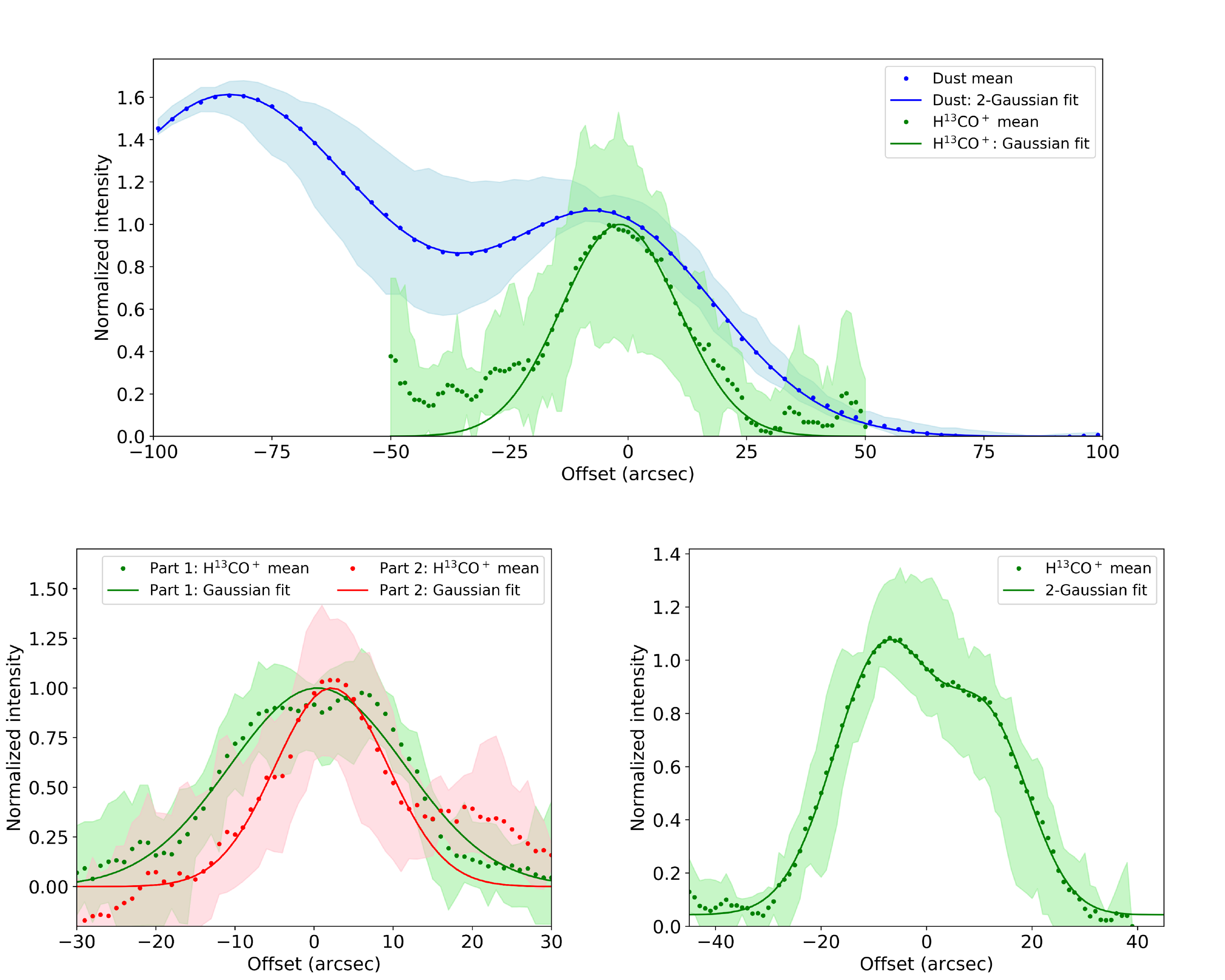}
\caption{Filament width in different regions. The dark dots represent the normalized average of the cuts over a filament section, which is fitted with one or two Gaussians. The light color spread represents the range of normalized values over a section of the filament. \textit{Top}: Serpens South NW filament --  Comparison of filament widths in dust (blue) and \htco\ (green). The dust map shows two parallel filaments about 75\arcsec\ apart and is fitted with two Gaussians. The FWHM of the filament mapped by us is 53$\pm$3\arcsec\ in dust and 29$\pm$5\arcsec\ in \htco.  \textit{Lower-left}: Serpens Main E filament -- Comparison of filament widths in two parts of the same filament -- FWHM 26$\pm$5\arcsec\ near cloud center (green) and 17$\pm$5\arcsec\ away from cloud center (red). Closer to the cloud core, the intensity profile across the filament departs from a Gaussian profile, even though we detect a single velocity-coherent component throughout. \textit{Lower-right}: Northern part of Serpens Main S filament -- Two parallel sub-filaments separated by about 25\arcsec\ with FWHM 23$\pm$8\arcsec\ and 19$\pm$7\arcsec\ respectively. A 2-Gaussian fit is used since two velocity-coherent sub-filaments were identified using the data cube.}
\label{fig:FilWidth}
\end{figure*}

In many of the regions, the presence of overlapping sub-filaments prevents us from determining their individual widths by this method, both in \htco\ and dust maps. We apply this method to isolated filaments (Serpens South - NW filament and Serpens Main - E filament) and sections of filaments not having any overlapping sub-structure. We use two-Gaussian fits for regions where partially overlapping parallel sub-filaments have ridges separated by more than twice the beam size (see Figure \ref{fig:FilWidth} lower-right). The Serpens South - NW filament has a deconvolved FWHM width of 0.059 $\pm$ 0.011 $pc$ in \htco\ and 0.083 $\pm$ 0.006 $pc$ in dust, while the corresponding values for Serpens Main - E filament are 0.043 $\pm$ 0.011 $pc$ and 0.067 $\pm$ 0.026 $pc$ respectively.  Our analysis shows that even in the isolated filaments, \htco\ maps do not have a smooth Gaussian profile (see Figure \ref{fig:FilWidth} lower-left). The widths vary along the length of the filaments by up to a factor of two. In regions where we could compare the dust and \htco\ widths for isolated structures, we found that the deconvolved widths are lesser in the \htco\ maps than the dust maps by a factor of about 1.5 (see Figure \ref{fig:FilWidth} top). Overall, we report filament deconvolved FWHM widths in the range 0.03 $pc$ to 0.08 $pc$ using \htco\ which is comparable to the \nnh\ widths for these filaments reported in \cite{Lee2014} and \cite{FL2014}. Our width estimation from the channels maps gives similar values, although there we estimated the widths of the overlapping structures as well that could not be determined using the integrated intensity maps.

The CLASSy-II results show that the widths of the filaments in dense gas tracers are on average significantly less than the \textit{Herschel} filament widths. The greater widths in the \textit{Herschel} maps cannot be solely explained by the presence of unresolved sub-structures, because we find this to hold true for isolated structures as well. This could indicate that the dust emission is tracing material which is intrinsically more extended than the distribution of dense gas in filaments.

\subsubsection{Multiple structures}

Except for the Serpens South - NW filament, all the remaining regions have partially overlapping multiple structures in the line of sight. The regions in Serpens Main have a line of sight velocity difference of as much as 1.4 $km/s$. Assuming timescales comparable to the cloud free-fall times ($\sim$ 1 $Myr$) for motion governed by gravity (which also includes turbulence in a bound molecular cloud), these structures should be separated by about 1.5 $pc$. This is comparable to the size of the molecular clouds and is much larger than the filament widths. Alternatively, if they are assumed to be in closer physical proximity, then they have proportionately lesser free-fall times, and therefore represent transient structures that may be forming from or evolving into larger structures. This discussion does not take into account effects of magnetic fields which can also affect the timescales.

These velocity-coherent structures are also distinct from each other in their morphology, velocity gradients and HNC absorption characteristics. Many of the sub-filaments are parallel to each other. Additionally in Serpens South - E region and NGC 1333 - SE region, the parallel sub-filaments have a fairly constant velocity difference of about 0.5 $km/s$ at the nearer edge between the filaments. In both these regions, there is a velocity gradient across one of the filaments. 

\cite{Hacar2013} also observed multiple velocity components in the L1495/B213 filaments in Taurus. They identified shorter 0.5 $pc$ coherent non-interacting sub-filaments which have velocity separations of 0.5-1.0 $km/s$ similar to what we observe. \cite{Taf2015} proposed a `fray and fragment' model to explain the multiple structures. This model starts with a wide filament that fragments into sub-filaments, which then further fragment into cores. Hydrodynamic simulations of turbulent clouds by \cite{Moeckel2015} and \cite{Smith2016} also showed the presence of multiple components in filaments. However contrary to the `fray and fragment' model, \cite{Smith2016} proposed a `fray and gather' model, in which the sub-filaments are formed first and then gathered together by large-scale motions within the cloud -- initially by large scale turbulent modes and afterwards gravitationally. Since the sub-filaments observed by us are parallel, they are likely to be influenced by the same physical processes locally, but the observations cannot distinguish between the two models discussed above.

\subsubsection{Absorption features in HNC}
\label{sec:Abs}

The HNC spectrum shows absorption features in many regions. These dips result in the HNC spectrum having multiple peaks, and are identified as absorption features based on the peaks in the \htco\ spectrum and the \nnh\ isolated hyperfine spectrum. In many regions, the dips correspond to similar dips in \hco\ and HCN. In regions having multiple structure, we see that in some cases only one structure has absorption dips, while in other cases emission from both structures have absorption dips.

\begin{figure*}
\centering
\includegraphics[width=0.7\textwidth]{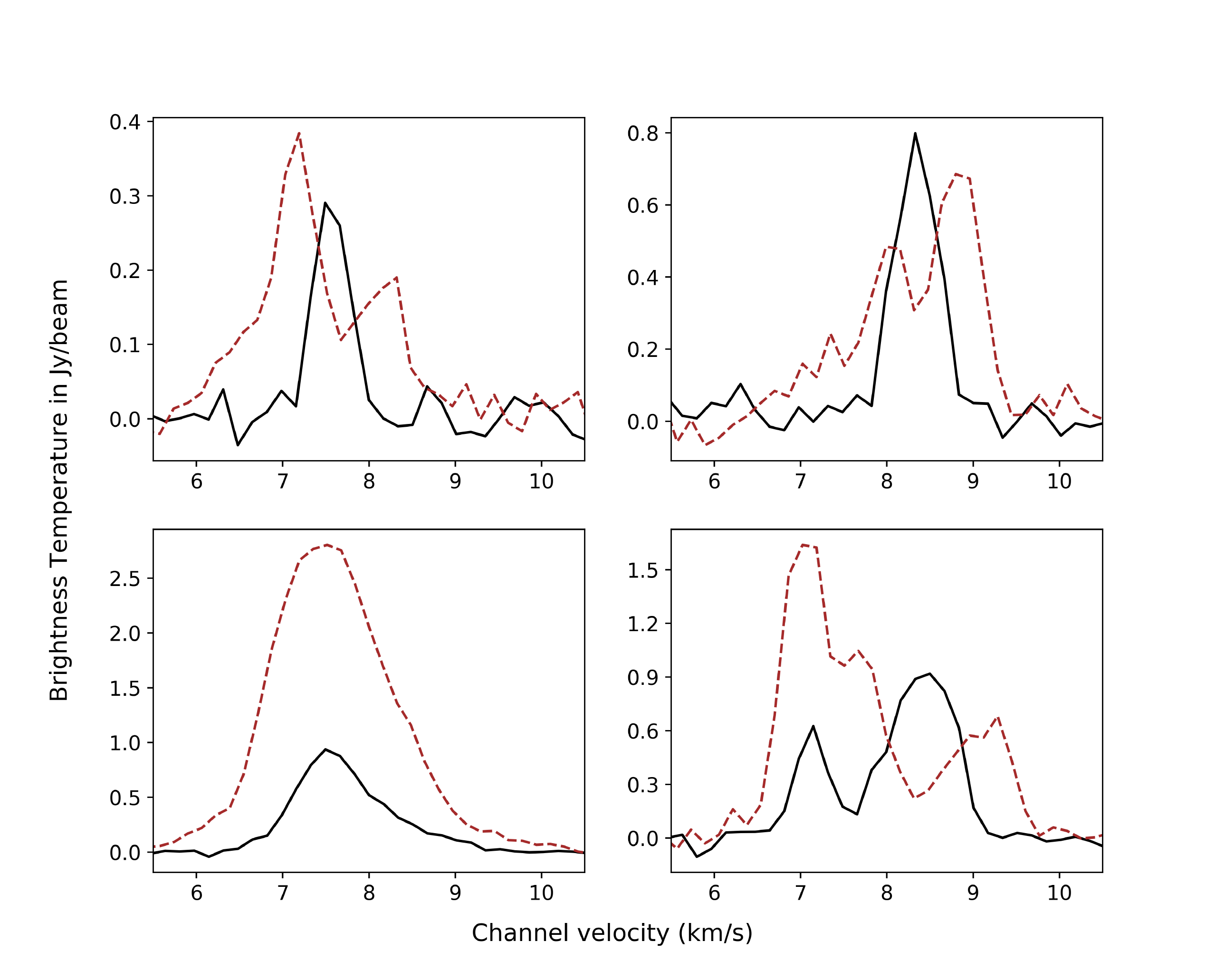} 
\caption{\htco\ (black, solid) and HNC (brown, dashed) spectra in various regions -- \textit{Upper-left}: Example of single velocity component with HNC absorption profile having higher blue-peak. This is found along the Serpens South - NW filament. \textit{Upper-right}: Example of single velocity component with HNC absorption profile having higher red-peak. This is observed in Serpens South - E region and in sub-filament `B' of Serpens Main - S region. \textit{Lower-left}: Example of single velocity component with no HNC absorption. This is observed in the Serpens Main cloud center. \textit{Lower-right}: Example of two velocity components one of which shows absorption in HNC, while the other does not. This is found in many regions in Serpens Main and in NGC 1333 - SE region. Additionally there are regions having two or more components, more than one of which show absorption (not shown in figure).}
\label{fig:AbsSpec}
\end{figure*}

Absorption features with a higher blue-shifted peak and a lower red-shifted peak are considered as a signature of radially symmetric infall into the filament core \citep{DVM2005, Friesen2013}. We find that in the regions studied by us, there are absorption features with both blue-asymmetry and red-asymmetry (see Figure \ref{fig:AbsSpec}). The generic infall models are inadequate in explaining the velocity structure of the HNC lines. 

Although self-absorption within the filament is likely, an alternate possibility is absorption by lower density clouds surrounding the main filament that are in the line-of-sight. This theory is supported by the observation that the absorption features in filaments are equally strong as we move from the center of the filament to the edges. Self-absorption by filament material should decrease through lesser optical depth regions near the edges of the filament.

\subsubsection{Filaments in relation to star formation}

Filaments are known to be closely associated with star forming regions, and many YSOs and prestellar cores are identified along some filaments \citep{Bontemps2010, Hacar2011}. On comparing our regions with the \textit{Herschel} 70\micron\ detections and the \textit{Spitzer} catalog of YSOs \citep{Cat2015}, we find that two of the filaments harbor multiple YSOs along their length, --the Serpens South - E filament (left panel of Figure \ref{fig:AnaSERPSE}) and the Serpens Main - S filament (left panel of Figure \ref{fig:AnaSERPMS}). The YSOs in these filaments are all Class 0/I/Flat sources indicating that they are associated with early stages of star formation. Both these filaments have parallel sub-structures and it cannot be determined whether some of the YSOs are associated with one sub-filament or the other, since they appear in the overlapping regions. YSOs are also detected at the filament intersection in the Serpens Main - S region and the filament-flow intersection in the Serpens Main cloud center (left panel of \ref{fig:AnaSERPME}), corroborating with observations by \cite{JS2014} towards the IRDC G035.39–00.33. This suggests that the sub-filaments may be interacting with each other, and are probably in close proximity even in the line-of-sight. We also find that all of the regions have continuum sources close to the ends of the filaments near filamentary hubs or cloud centers.

Different filaments can be at different stages of their star forming life \citep{Myers2017}, and based on our observations it can be argued that only two of the five filaments studied by us are currently actively star forming. The mass per unit length of filaments is often used as an indicator of the evolutionary stage of filament accretion \citep{Hei2013, Palm2013, Li2014}, with higher values indicating greater chances of gravitational fragmentation. We can estimate the mass per unit length of the \textit{Herschel} filaments by summing over the pixels of the H$_2$ column density map over the filament and subtracting the background. On applying this method to the isolated filaments, we obtain an average mass per unit length of 21.3 $\Msun/pc$ for Serpens South - NW filament and 16.9 $\Msun/pc$ for Serpens Main - E filament. The values are comparable to or lesser than the critical mass per unit length of 20.9 $\Msun/pc$, calculated for isothermal (T = 12.5 $K$) self-gravitating cylinders using the formula $ M_{L,crit} = 2c_s^2 /G$ where $c_s$ is the sonic speed in the cloud and $G$ is the universal Gravitational constant \citep{Ost1964}. This is consistent with the observation that neither of these filaments have any Class 0/I sources along their lengths. The Serpens South - E filament and the Serpens Main - S filament, which have YSOs along their length, have mass per unit length values of 28.7 $\Msun/pc$ and 44.6 $\Msun/pc$ respectively, which are both greater than the critical value. However the 36\arcsec\ column density map beam-size is insufficient to resolve the contribution of individual sub-structures within the filament. In the NGC 1333 - SE filament, the mass per unit length varies by a factor of 4; although its mean value is super-critical, it does not have any YSOs along its length. So even though the mass per unit length is a good indicator of the star formation stage for a filament, it is not a conclusive discriminator.

\subsection{Kinematics}

All the CLASSy-II regions have at least one filamentary structure with an evident gradient in the line-of-sight velocity: --across the filament, along the filament or both. The complete list is given in Table \ref{tbl:Grad}. The filaments and regions where we observe negligible velocity variations may still have gradients into the plane of the sky.

The kinematics observed in \nnh\ match well with that in \htco\ for the non-overlapping regions, and show similar trends in regions having multiple structures in the line-of-sight. This is evident from the velocity maps in Section \ref{sec:Res}. However, the hyperfine structure of \nnh\ limits its capability of distinguishing between multiple structures and quantifying their kinematic features independently. \nnh\ velocity maps had been generated using line-fitting of all the 7 hyperfine components, but assuming a single velocity component for most locations on the map. In regions with multiple velocity components, such a line-fitting produced a centroid velocity with a large line-width. Velocity maps thus obtained were used previously to determine the kinematics of the regions \citep{Lee2014, FL2014}, which in some cases give different results from our \htco\ analysis that allows for multiple components. Attempts to fit for multiple velocity components in the \nnh\ spectra leads to erroneous degenerate solutions, unless the components are $\geqslant$ 1 $km/s$ apart. For example, in \cite{FL2014} the Serpens South - E filament is reported to have a large velocity gradient of 11.9 $km\ s^{-1} pc^{-1}$ across the filament. However, the \htco\ data cube reveals that this is an effect of multiple velocity-coherent components in close proximity (see Figure \ref{fig:AnaSERPSE}), even though its velocity map matches well that of \nnh\ (see Figure \ref{fig:SERPSE}). Only the eastern half of one of the sub-structures (`A') in this filament has a gradient across it, but in the opposite sense to that reported in \cite{FL2014}.

\subsubsection{Velocity gradients across filaments and its implications on filament formation mechanism}

Four of the eight identified filamentary structures have velocity gradients perpendicular to their length in the range $3.6-12.2$ $km\ s^{-1} pc^{-1}$. They are determined from the \htco\ maps as discussed in Section \ref{sec:Analysis}. These gradients are unidirectional in each filamentary structure but have variations in magnitude along the filaments. The magnitudes correspond to crossing times of $1-3$ $\times$ $10^5$ years. Such gradients of smaller magnitude have been observed by \cite{Per2014} as well. \cite{Sch2010} also reported observations of gradients across a filament in the Cygnus molecular cloud but it has a width of 1 $pc$, about 20 times wider than the filaments observed by us. Further, the gradient across the Cygnus filament changes direction at different positions along the length of the filament. Out of the regions studied by us, in addition to the four filamentary structures with gradients across their widths, we also identified line-of-sight velocity gradients across the Serpens Main - S region, but it is arguably an effect of multiple juxtaposed structures at different velocities.

\begin{figure*}
\centering
\includegraphics[width=0.5\textwidth]{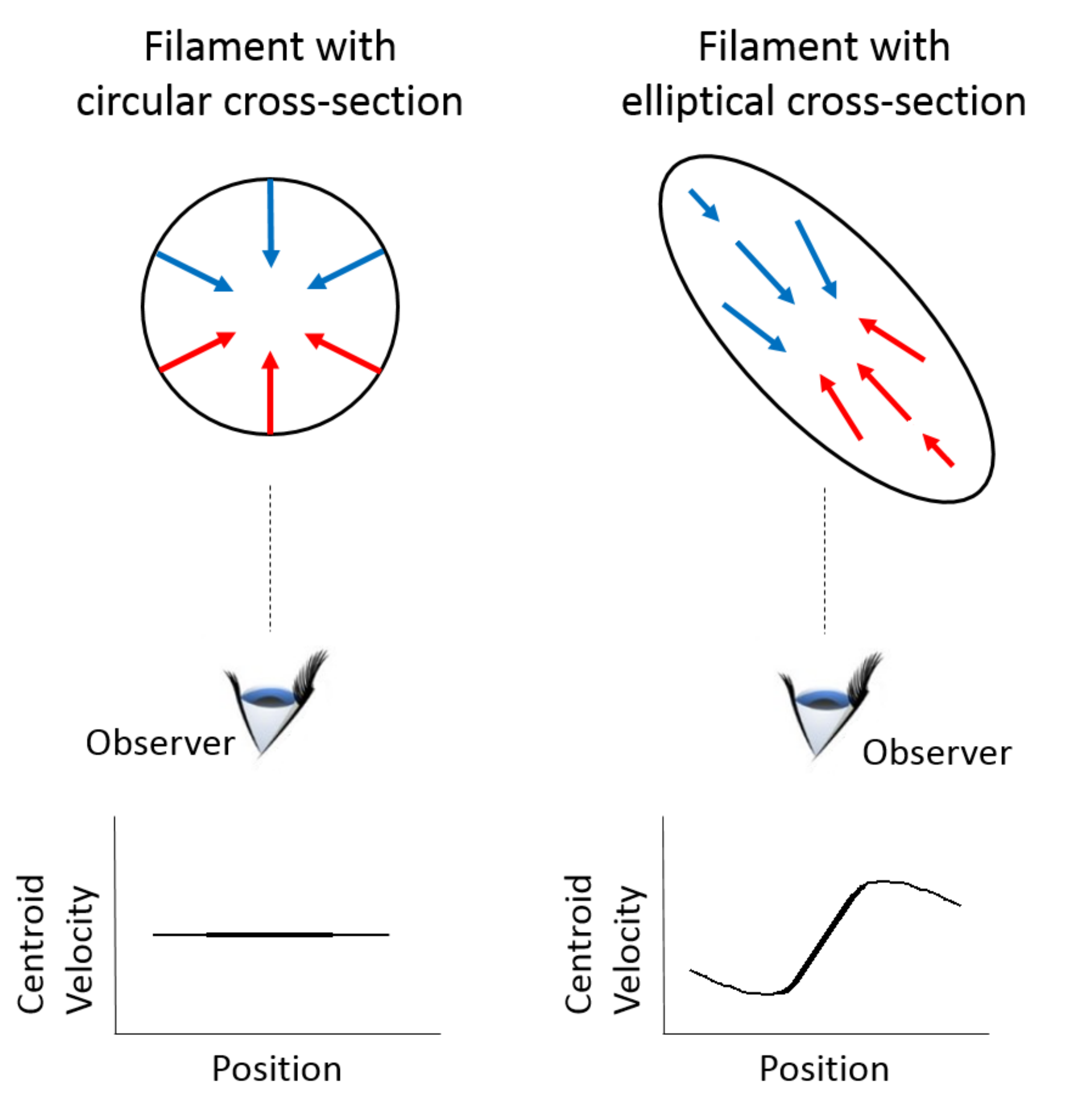} 
\caption{Simplified schematic of two types of filament cross-sections -- circular and elliptical, indicating that gas accretion towards the filament core has different kinematic signatures in the two cases. In the isotropic accretion case (left), the centroid velocity remains constant for a cut across the filament. In the non-axisymmetric case (right), depending on how the filament cross-section is oriented with respect to the sky plane, it can result in a gradient in the centroid velocity vs position plot. The blue-shifts and red-shifts are expected to be greater closer to the filament core compared to the edges. This causes a change in the slope going from the center to the edges, though it may not be evident in the observations since the emission is also lesser in these parts. The plots are simplified and do not show effects of velocity dispersion which is expected to increase the velocity spread near the filament core in both cases.}
\label{fig:Across}
\end{figure*}

The velocity gradients across the filaments support the filament formation model by Chen and Ostriker (private communication). According to the model, the velocity gradient is a projection effect of the accreting material in a 2-D flow within the dense layer created by colliding turbulent cells. The cartoon in Figure \ref{fig:Across} illustrates this effect. In case of a non-axisymmetric cylinder in the sky plane, we expect the observed centroid velocity to vary systematically for a cut across the filament. This model also corroborates with \cite{Smith2016}, who report that the filament cross-sections in their simulations are elongated instead of being circular, and that the largest gradients appear perpendicular to the filament. The absence of gradients across some of the filaments could be a result of close to face-on viewing angle or a different formation mechanism. More observations are required to establish the broad relevance of this model. 

Alternate interpretations of the velocity gradient include filament rotation \citep{OT2002} and multiple narrow filaments that are partially overlapping in the sky viewing plane \citep{Beu2015}. To our knowledge, the first idea is not supported by numerical simulations \citep{Smith2016}. We have seen some evidence of parallel sub-filaments in a few regions masquerading as a single filament with a large gradient if we only see their velocity maps (as in Serpens South - E region and NGC 1333 - SE region). However after disentangling their individual velocity distributions using the \htco\ maps, we see that one of the parallel sub-filaments has a velocity gradient across it independent of the other sub-filament (see Figure \ref{fig:AnaN1333SE} lower-right). The observed gradients across filaments (which are equally or more evident than gradients along filaments) indicate that the local dynamical evolution of these filaments occur much faster than the rate of flow of material along them.

\subsubsection{Velocity gradients along filaments and its implications}
\label{sec:VelAlong}

Two filamentary structures in Serpens Main have velocity gradients along their lengths of 2.6 and 4.6 $km\ s^{-1}pc^{-1}$ in \htco\ (Figures \ref{fig:SERPME} and \ref{fig:AnaSERPMS}). They both have one end close to the cloud core, and another end away from it. Additionally there are distinct velocity gradients along multiple flows in the hub region (that are not long enough to be classified as filamentary structures). As shown in Figure \ref{fig:AnaSERPME}, the flows intersect at the same point along with one of the filamentary structures. The gradients are maximum closer to the cloud core increasing by a factor of $\sim$ 4 compared to the gradients near the other end of the filamentary structures. This may indicate accelerated motions near regions of higher gravitational potential as the structures fall into them. 

The different interpretations of this gradient discussed in \cite{JS2014} include cloud rotation, unresolved sub-filament structures, accretion along the filaments and global gravitational collapse. In the Serpens Main - South region, although cloud rotation is unlikely based on the large scale kinematic signatures of the region, the gradient could be caused by any of the other scenarios, or due to projection effects. In the Serpens Main cloud center, the presence of multiple converging structures with accelerated gradients along their lengths is suggestive of global gravitational collapse \citep{Peretto2013}. In this scenario, matter from large scales are gathered to the center of a gravitational potential well, where multiple protostars are formed. It is to be noted that material does not flow along the filaments (analogous to water flowing in a river), rather the entire filament moves down into the potential well.

\begin{figure*}
\centering
\includegraphics[width=0.5\textwidth]{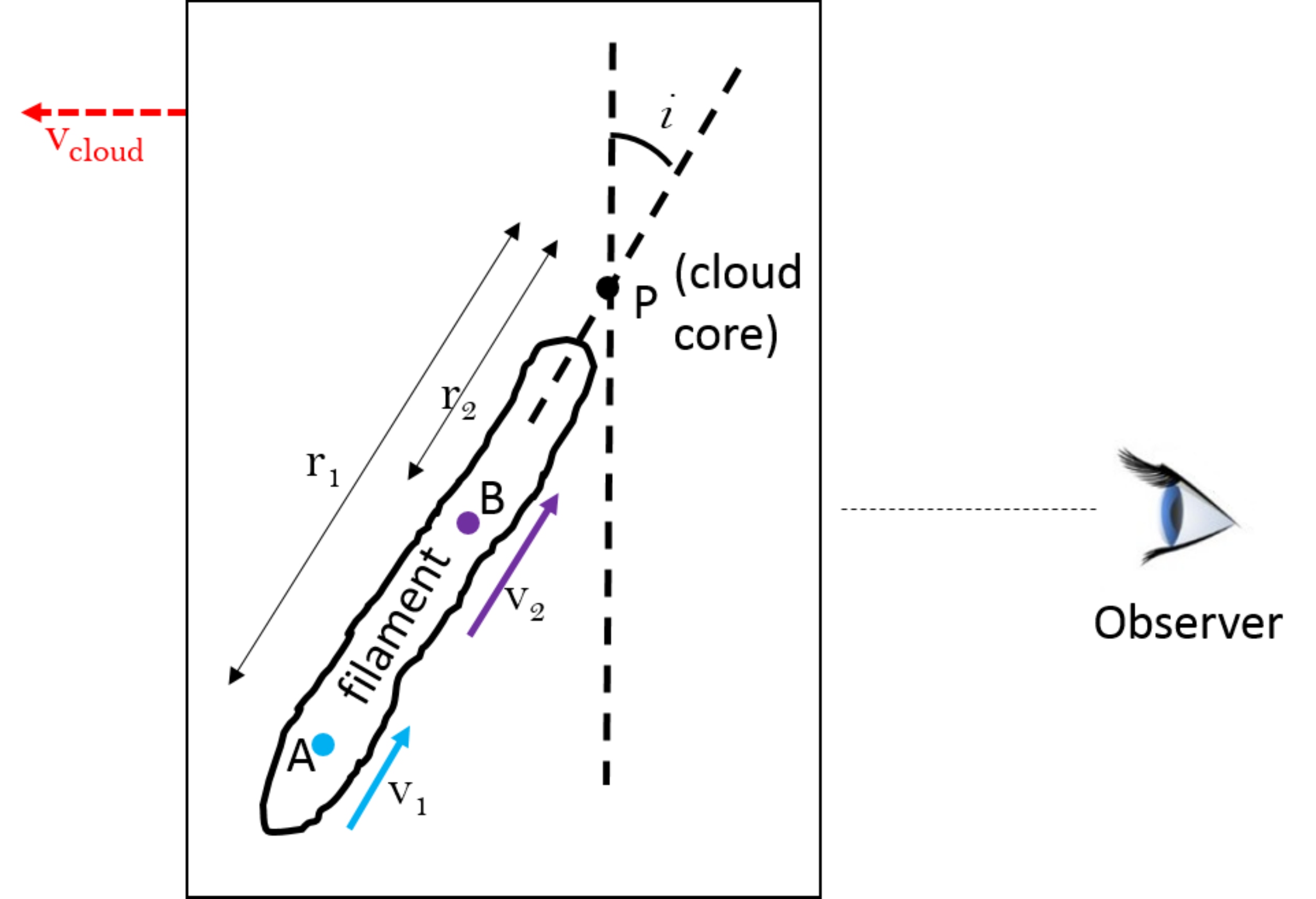} 
\caption{Simplified schematic of a filament at an inclination angle with respect to the sky plane. The filament is assumed to be accelerating along its length due to the gravitational field of a massive core at P. B is closer to the core and has a greater velocity along the filament than A. In this configuration with the filament inclined away from the observer, emission from B is more blue-shifted (or less red-shifted) compared to emission from A.}
\label{fig:Along}
\end{figure*}

In the Serpens Main cloud center, we identify a filament intersection point having the highest blue shift. We use velocities ($v_{obs}$) at different distances ($l_{obs}$) from this point to measure the gravitational potential. The line-of-sight cloud velocity ($v_{cloud}$) of 8.15 $km/s$ is subtracted from the observed line-of-sight velocities. The velocity thus obtained is a projection of the velocity along the filament $v$ in the local frame of the cloud (see Figure \ref{fig:Along}). We obtain $v$ assuming an inclination angle $i$ (positive when filament inclined away from observer), using the expression $v_{obs} = v_{cloud} - v\sin{i}$. The observed distance $l_{obs}$ is also a projection of the radial distance $r$ and they are related as $l_{obs} = r\cos{i}$. Assuming a steady state, using energy conservation for a pair of points along the filament, we can write
\begin{equation}
{v_2}^2 - {v_1}^2 = GM\Big(\frac{1}{r_2}-\frac{1}{r_1}\Big)
\end{equation}
where $G$ is the Gravitational Constant and $M$ is the mass of the core. This can be written in terms of the observables and the inclination angle as 
\begin{equation}
(v_{cloud}-v_{2,obs})^2 - (v_{cloud}-v_{1,obs})^2 = GM \sin^2{i} \cos{i}\Big(\frac{1}{l_{2,obs}}-\frac{1}{l_{1,obs}}\Big)
\end{equation}

Using this equation, and measured sets of velocities at different distances from the core for the different structures, we obtain consistent core mass values in the range $30-37$ $\Msun$ for an inclination angle of $i$ = 45\deg. The mass varies by a factor of about 1.5 if we assume an inclination angle range of $30-60$ degrees. Additionally, because of uncertainties in the distance and velocity measurements, there can be up to 50\% error in mass estimates. As shown in the figure, from this interpretation, we can also make inferences about the 3-D structure of the filaments --whether the filaments are inclined towards us or away from us. For both these cases, the more blue-shifted (or less red-shifted) part of the filament is expected to be closer to the observer than the less blue-shifted (or more red-shifted) part.

%% file: summary.tex
\section{Summary}
\label{sec:Sum}

We presented CARMA observations of \htco(1-0) and HNC(1-0) for five regions in Serpens Main, Serpens South and NGC 1333 containing filaments. For the four Serpens regions, we also obtained data on the \htcn(1-0) emission. The observations have an angular resolution of $\sim$ 7\arcsec\ and a spectral resolution of 0.16 $km/s$. We studied the morphology and kinematics of these regions, comparing them to existing maps of the dust continuum and \nnh(1-0) emission. Our main conclusions are summarized below.
\begin{itemize}
\item The emission distribution in the \htco\ maps traces similar filamentary structures as the \nnh\ maps obtained by CLASSy, and correspond to the same morphology and kinematics.
\item In many regions, multiple velocity-coherent structures are present, identifiable by multiple peaks in the \htco\ spectrum. \htco\ is the only species observed by us that allows us to unambiguously disentangle the overlapping components.
\item Some of these multiple structures are filament sub-structures that are roughly aligned with each other even though they have velocity differences of $0.5-1.0$ $km/s$ between them. We identify 2 sub-structures each in 3 filaments, while 2 filaments are each found to be comprised of a single velocity-coherent component. We report statistics of these 8 filamentary structures.
\item The mean width of these filamentary structures is 0.05 $pc$, but they vary in the range of $0.03-0.08$ $pc$. Along the same filamentary structure, the width can vary by a factor of 2. The widths of velocity-coherent filaments in the dense gas tracers are a factor of 1.5 narrower than the \textit{Herschel} widths.
\item Four of the eight filamentary structures have significant velocity gradients perpendicular to the filament length, with mean values in the range of $4-10$ $km\ s^{-1} pc^{-1}$. This provides evidence for predictions from simulations, in which filaments form via inflows within the dense layer created by colliding turbulent cells.
\item Two filamentary structures in Serpens Main have velocity gradients along their lengths, which increase closer to the cloud core. This may indicate gravitational inflow of filaments into the central core.
\item Class 0/I/Flat YSOs are identified only along two of the filaments, and are found to be preferably located at the overlapping regions of the filamentary structures. This suggests that the sub-filaments are physically interacting with each other, which possibly plays a role in star formation.
\end{itemize}

Overall, the observations support the presence of finer structures within the \textit{Herschel} filaments with systematic properties like alignment of sub-filaments and presence of velocity gradients across them. These properties suggest a common formation mechanism. Features like the large distribution of widths of filamentary structures and the presence of YSOs only in some of the regions indicate their diversity. These can arise from local effects or could be dependent on the evolutionary stage of the filaments. 